\documentstyle[psfig]{l-aa}

\newcommand{\reduceme}{\mbox{R\raisebox{-0.35ex}{E}D\hspace{-0.05em}%
\raisebox{0.85ex}{uc}\hspace{-0.90em}\raisebox{-.35ex}{{m}}\hspace{0.05em}E}}

\begin{document}

\thesaurus{ 08.01.2, 08.02.1, 08.03.3, 08.12.1, 08.09.2 }

\title {Multiwavelength optical observations of chromospherically 
active binary systems}
\subtitle{II. EZ Pegasi
\thanks{Based on 
observations made with the Isaac Newton, William Herschel 
and Jacobus Kapteyn telescopes
operated on the island of La Palma by the
Royal Greenwich Observatory at
the Spanish Observatorio del Roque de Los Muchachos of the
 Instituto de Astrof\'{\i}sica de Canarias
} }

\author{
D.~Montes
\and J.~Sanz-Forcada
\and M.J.~Fern\'{a}ndez-Figueroa
\and E.~De Castro
\and A.~Poncet  
}

\offprints{ D.~Montes}

\institute{Departamento de Astrof\'{\i}sica,
Facultad de F\'{\i}sicas,
 Universidad Complutense de Madrid, E-28040 Madrid, Spain
\\  E-mail: dmg@astrax.fis.ucm.es}

\date{Received ; accepted }

\maketitle

\begin{abstract}


The star EZ Peg, long ago classified as cataclysmic variable, has been shown
to be a chromospherically active binary system of the RS CVn-type.
In this paper we have analysed, using the spectral subtraction technique,
simultaneous spectroscopic observations of the H$\alpha$,  H$\beta$, 
Na~{\sc i} D$_{1}$ and D$_{2}$, He~{\sc i} D$_{3}$, Mg~{\sc i} b triplet,
Ca~{\sc ii} H \& K, and Ca~{\sc ii} infrared triplet lines.
We have found that the hot component is the active star of the system, 
showing strong emission in the H$\alpha$, Ca~{\sc ii} H \& K, H$\epsilon$,
and  Ca~{\sc ii} IRT lines, 
and a strong filling-in of the H$\beta$ line,
however the Na~{\sc i} D$_{1}$ and D$_{2}$ and Mg~{\sc i} b triplet lines
do not present filled-in.
The He~{\sc i} D$_{3}$ could present a total filling-in due to microflaring. 
The observed variations (in different epochs and with the orbital phase)  
of the different activity indicators, 
formed at different height in the chromosphere,
are correlated.
Very broad wings have been found in the subtracted profiles 
of H$\alpha$ and Ca~{\sc ii} IRT $\lambda$8498 and $\lambda$8662 lines.
These profiles are well matched using a two-component
Gaussian fit (narrow and broad) and the broad component
could be interpreted as arising from microflaring.
The higher luminosity class of the hot component,
 that our spectra seem to indicate, 
could explain why the hot component is the active star of the system.

%
%

\end{abstract}


\section{Introduction}

EZ Peg (BD +24$^{\circ}$ 4742)  has a misleading history of confused 
interpretation of the origin and composition of the system. 
The spectrum of EZ Peg was 
originally classified by Cannon as G5 (Schlesinger et al. 1934). 
Vyssotsky \& Balz (see Alden 1958) claimed that the star varies 
from B0 to G5, due to a spectrum taken in 1943 
which appeared to be a B star spectrum.
 Later Howell \& Bopp (1985) presented evidence that this erroneous 
interpretation 
of the spectrum was due to poor resolution data. 
Irvine (1972) proposed that EZ Peg was an U Gem system (a cataclysmic 
variable), but it is an unlikely explanation due to the normal UV emission of 
the system: in a color-color plot there is no strong UV excess typical of the 
dwarf novae and symbiotics (Szkody 1977). Furthermore linear polarimetry 
measurements of the 
system (Szkody et al. 1982) show that EZ Peg does not seem to be a cataclysmic
variable. 

Howell \& Bopp (1985) observed the double-lined spectrum 
of EZ Peg and interpreted the color-indices (B-V) and (U-B) as
that of a composite spectral type of G5 IV + K0 IV. 
They observed Ca~{\sc ii} H \& K, 
H$\alpha$ and H$\epsilon$ lines in emission, and they supposed that this 
emission originated in the cooler star, because it is the most typical case.
Howell et al. (1986) suggested that both components of EZ Peg were subgiants,
and pointed out a slightly asymmetric profile in H$\alpha$, interpreted as
a vertical motion in the atmosphere of the active star of the system.

Griffin (1985) presented radial-velocity observations 
of the system, and he calculated the orbital elements. 
Recently Gunn et al. (1996) made cross-correlation radial velocity 
measurements of 
EZ Peg, showing a disagreement with the phases calculated from the ephemeris 
given by Griffin (1985). 
Vilkki et al. (1986) calculated a distance to EZ Peg of 83~pc using the
trigonometric parallax method, and they suspected that EZ Peg belongs to the
Hyades cluster. Barret (1996) calculated a distance of 82~pc using 
linear polarimetry.

Finally, recent studies in the IR band by Mitrou et al. (1996) with the IRAS 
satellite indicated that the system does not seem to present any circumstellar 
matter. 

Continuing with the series of multiwavelength optical observations of 
chromospherically active binaries (Montes et al. 1997b, hereafter Paper I), 
in this paper we present simultaneous observations in different  
optical wavelength bands, including also a high resolution echelle spectrum.
All the data seem to point out that EZ Peg is a RS CVn system 
with an evolved G5 active star, 
and a K0V or IV non-active star. 
We have applied the spectral
subtraction technique, observing the Ca~{\sc ii} H \& K,
H$\epsilon$, H$\alpha$ and Ca IR triplet lines in emission, and a filling-in
of the H$\beta$ line, proceeding from the hotter component of the system.

In Sect.~2 we give the details of our observations and data reduction.
In Sect.~3 we discuss some of the stellar parameters of the binary system
 and in Sect.~4  the behavior of the different chromospheric 
activity indicators is described.
Finally in Sect.~5 we present the conclusions.

\section{Observations and Data Reduction}

Spectroscopic observations of EZ Peg in 
several optical chromospheric activity indicators 
have been obtained during two observing runs with 
the 2.5~m Isaac Newton Telescope (INT) at the
Observatorio del Roque de Los Muchachos (La Palma, Spain)  using the
Intermediate Dispersion Spectrograph (IDS).
In the first run (September 13-15 1995) we have obtained two spectra: in the 
H$\alpha$ and Na~{\sc i} D$_{1}$, D$_{2}$, He~{\sc i} D$_{3}$ line regions, 
using grating H1800V, camera 500
and a 1024~x~1024 pixel TEK3 CCD as detector.
The reciprocal  dispersion  achieved  is 0.24~\AA/pixel
which yields a spectral resolution of 0.48~\AA$\ $ and a useful wavelength
range of 250~\AA$\ $ centered at 6563~\AA$\ $ (H$\alpha$)
and 5876~\AA$\ $ (He{\sc i} D$_{3}$) respectively.
In the second run (November 26-28 1996) we used the same configuration 
but two different gratings 
H1200R and H1200B, for the red and blue spectral regions respectively.
Using the same configuration a reciprocal dispersion of 0.39~\AA$\ $
was obtained which yields a spectral resolution of 0.94~\AA$\ $ and a useful 
wavelength range of 385~\AA$\ $ centered at 
6563~\AA$\ $ (H$\alpha$),
5876~\AA$\ $ (He{\sc i} D$_{3}$), 
4861~\AA$\ $ (H$\beta$),
3950~\AA$\ $ (Ca~{\sc ii} H\&K) respectively.

In Table~\ref{tab:obslog} we give the observing log. 
For each spectral region observed we list
the date, UT, orbital phase ($\varphi$) and signal to noise ratio (S/N)
obtained.

We also analyse here echelle spectra 
obtained with the 4.2~m William Herschel Telescope (WHT) and the
Utrecht Echelle Spectrograph (UES) on July 31 1993 that we have 
retrieved from La Palma Data Archive (Zuiderwijk et al. 1994).
These WHT/UES spectra were obtained with echelle 31
(31.6 grooves per mm) and a 1280~x~1180 pixel EEV6 CCD as detector.
The central wavelength is 6127~\AA$\ $ and covers a wavelength range
from 4845 to 8805~\AA$\ $ over 53 echelle orders.
The achieved reciprocal dispersion ranges from 0.045 to
0.080~\AA/pixel.
In order to improve the S/N ratio and to avoid cosmic ray events, 
the final spectrum analysed here is the resulting from  
combine 28 CCD images with exposure times of 240~s each, obtaining
a final spectrum with a S/N of 315 in the H$\alpha$ line region.  
In Table~\ref{tab:uesjul93} we give the wavelength range and the 
spectral lines of interest in each echelle order. 

In addition we also use one spectrum in the region of the 
Ca~{\sc ii} IRT lines taken on January 16 1997
with the 1.0~m Jacobus Kapteyn Telescope (JKT) using the 
Richardson-Brealey Spectrograph (RBS) with a grating of 
1200 grooves per mm, and a 1024~x~1024 pixel TEK4 CCD as detector.
The reciprocal  dispersion  achieved is 0.85~\AA/pixel
which yields a spectral resolution of 1.7~\AA$\ $ and a useful wavelength
range of 865~\AA$\ $ centered at 8765~\AA.

The spectra have been extracted using the standard
reduction procedures in the IRAF
\footnote{IRAF is distributed by the National Optical Observatory,
which is operated by the Association of Universities for Research in
Astronomy, Inc., under contract with the National Science Foundation.}
 package (bias subtraction,
flat-field division, and optimal extraction of the spectra).
The reduction of the JKT/RBS spectrum  was carried out 
using the reduction package \reduceme\ (Cardiel \& Gorgas 1997, 
see also {\small http://www.ucm.es/OTROS/Astrof/reduceme/reduceme.html}). 
The wavelength calibration was obtained by taking
spectra of a Cu-Ar lamp in the INT/IDS spectra, a Th-Ar lamp in the 
WHT/UES spectrum and a Cu-Ne lamp in the JKT/RBS spectrum.
Finally, the spectra have been normalized by 
a polynomial fit to the observed continuum.

\begin{table*}
\caption[]{Observing log
\label{tab:obslog}}
\begin{flushleft}
\scriptsize
\begin{tabular}{cccccccccccccccccccc}
\noalign{\smallskip}
\hline
\noalign{\smallskip}
Date &
\multicolumn{15}{c}{IDS-INT} &\ &
\multicolumn{3}{c}{UES-WHT} \\
\cline{2-16}\cline{18-20}
\noalign{\smallskip}
     &
\multicolumn{3}{c}{H$\alpha$ } &\ &
\multicolumn{3}{c}{Na~I D$_{1}$, D$_{2}$, He~I D$_{3}$ } &\ &
\multicolumn{3}{c}{Ca~II H \& K } &\ & 
\multicolumn{3}{c}{H$\beta$ } &\ &
\multicolumn{3}{c}{Echelle} \\
\cline{2-4}\cline{6-8}\cline{10-12}\cline{14-16}\cline{18-20}
\noalign{\smallskip}
     &
\tiny
\tiny UT & {$\varphi$} & \tiny S/N &\ &
\tiny UT & {$\varphi$} & \tiny S/N &\ &
\tiny UT & {$\varphi$} & \tiny S/N &\ &
\tiny UT & {$\varphi$} & \tiny S/N &\ & 
\tiny UT & {$\varphi$} & \tiny S/N
\scriptsize
\\
\noalign{\smallskip}
\hline
\noalign{\smallskip}
%
 1993/07/31 &     - &     - &   - & &     - &     - &   - & & 
                  - &     - &   - & &     - &     - &   - & & 02:05 & 0.954
& 315  \\
\noalign{\smallskip}
 1995/09/14 & 00:36 & 0.416 & 295 & & 00:04 & 0.414 & 362 & & 
                  - &     - &   - & &     - &     - &   - & &  - & - \\
 1995/09/15 & 00:23 & 0.501 & 332 & & 00:02 & 0.500 & 343 & & 
                  - &     - &   - & &     - &     - &   - & &  - & - \\
\noalign{\smallskip}
 1996/11/26 & 21:25 & 0.141 & 295 & &     - &     - &   - & &
              23:11 & 0.148 & 295 & &     - &     - &   - & &  - & - \\
 1996/11/27 & 20:39 & 0.224 & 295 & & 20:49 & 0.225 & 362 & &
              19:35 & 0.220 & 295 & &     - &     - &   - & &  - & - \\
 1996/11/28 & 19:24 & 0.306 & 295 & & 20:25 & 0.310 & 362 & &
              21:44 & 0.314 & 295 & & 21:19 & 0.312 & 362 & &  - & - \\
\noalign{\smallskip}
\hline
\end{tabular}

\end{flushleft}
\end{table*}

\begin{table}
\caption[ ]{UES spectral orders 
\label{tab:uesjul93} }
\begin{flushleft}
\scriptsize
\begin{tabular}{l l l l l l l l l l l l l}
\hline
\noalign{\smallskip}
 No.     & $\lambda$$_{i}$  & $\lambda$$_{f}$  & Chromospheric Lines & 
Other lines \\
\noalign{\smallskip}
\hline
\noalign{\smallskip}
 1       & 4845.80 & 4897.29 & H$\beta$ & \\
 2       & 4887.53 & 4939.46 &          & \\
 3       & 4929.98 & 4982.35 &          & \\
 4       & 4973.16 & 5026.00 &          & \\
 5       & 5017.11 & 5070.41 &          & \\
 6       & 5061.85 & 5115.62 &          & \\
 7       & 5107.38 & 5161.63 &          & \\
 8       & 5153.74 & 5208.48 & Mg~{\sc i} b $\lambda$5167-72-83 & \\
 9       & 5200.94 & 5256.19 &          & \\
10       & 5249.02 & 5304.77 &          & \\
11       & 5297.99 & 5354.27 &          & \\
12       & 5347.89 & 5404.69 &          & Ti~{\sc i} $\lambda$5366\\
13       & 5398.73 & 5456.07 &          & Ti~{\sc i} $\lambda$5426 \\
14       & 5450.54 & 5508.43 &          & \\
15       & 5503.36 & 5561.81 &          & \\
16       & 5557.21 & 5616.23 &          & \\
17       & 5612.13 & 5671.73 &          & \\
18       & 5668.14 & 5728.34 &          & \\
19       & 5725.28 & 5786.09 &          & \\
20       & 5783.58 & 5845.01 &          & \\
21       & 5843.08 & 5905.14 & Na~{\sc i} D$_{1}$, D$_{2}$, He~{\sc i} D$_{3}$
& Ti~{\sc i} $\lambda$5866 \\
22       & 5903.82 & 5966.53 &          & \\
23       & 5965.84 & 6029.20 &          & \\
24       & 6029.17 & 6093.21 &          & \\
25       & 6093.87 & 6158.59 &          &  \\
26       & 6159.96 & 6225.39 &          &  \\
27       & 6227.51 & 6293.66 &          & \\
28       & 6296.55 & 6363.44 &          & \\
29       & 6367.15 & 6434.78 &  & Fe~{\sc i} $\lambda$6411.66 \\
30       & 6439.35 & 6507.75 &  & Fe~{\sc i} $\lambda$6495 \\
31       & 6513.20 & 6582.39 & H$\alpha$ & \\
32       & 6588.77 & 6658.76 &           & \\
33       & 6666.11 & 6736.93 &           & Fe~{\sc i} $\lambda$6678 \\ 
"        &         &         &           & Li~{\sc i} $\lambda$6708 \\
"        &         &         &           & Ca~{\sc i} $\lambda$6718 \\
34       & 6745.30 & 6816.95 &           & \\
35       & 6826.39 & 6899.90 &  & O$_{2}$ $\lambda$6867 \\
36       & 6909.45 & 6982.85 &          & \\
37       & 6994.56 & 7068.87 &          & \\
38       & 7081.80 & 7157.03 &          & \\
39       & 7171.24 & 7247.43 &          & \\
40       & 7262.98 & 7340.14 &          & \\
41       & 7357.09 & 7435.25 &          & \\
42       & 7453.68 & 7532.86 &          & \\
43       & 7552.84 & 7633.08 &  & B-band $\lambda$7600  \\
44       & 7654.68 & 7736.00 &  & K~{\sc i} $\lambda$7665, 7699  \\
45       & 7759.30 & 7841.73 &          & \\
46       & 7866.82 & 7950.40 &          & \\
47       & 7977.38 & 8062.12 &          & \\
48       & 8091.08 & 8177.03 &          & \\
49       & 8208.08 & 8295.27 &          & \\
50       & 8328.51 & 8416.99 &  & Ti~{\sc i} (M.33)  \\
51       & 8452.54 & 8542.33 & Ca~{\sc ii} IRT $\lambda$8498, 8542   
                                             &  Ti~{\sc i} $\lambda$8468 \\
"        &         &         &               &  Fe~{\sc i} $\lambda$8514 \\
"        &         &         &               &  Ti~{\sc i} $\lambda$8518 \\
52       & 8580.32 & 8671.46 & Ca~{\sc ii} IRT $\lambda$8662 &  
Fe~{\sc i} $\lambda$8621 \\
53       & 8712.02 & 8804.56 &          & \\
&        &        &          \\
\noalign{\smallskip}
\hline
\end{tabular}
\end{flushleft}
\end{table}

\begin{table*}
\caption[]{Stellar parameters
\label{tab:par}}
\begin{flushleft}
\scriptsize
\begin{tabular}{ c c c c c c c l l c }
\noalign{\smallskip}
\hline
\noalign{\smallskip}
{T$_{\rm sp}$} & {SB} & {R} &
 {d} & B-V & {V-R} & T$_{\rm conj}$ &{P$_{\rm orb}$} & {P$_{\rm rot}$} & Vsin{\it i}\\
               &      & (R$_\odot$) & (pc) &  &  & (H.J.D.) & (days) &
 (days) & (km s$^{-1}$)  \\
\noalign{\smallskip}
\hline
\noalign{\smallskip}
 G5V-IV/K0IV: & 2&  -  &  83  & 0.73/0.91 & 0.58/0.71 & 
2445737.165 &11.6598 & 11.6626 & 9/7 \\
%
%
\noalign{\smallskip}
\hline
\noalign{\smallskip}
\end{tabular}

\end{flushleft}
\end{table*}

\section{Stellar parameters of the binary system EZ Peg}

EZ Peg is a  double-lined spectroscopic binary 
with an orbital period of 11.6598 days (Griffin 1985), 
classified as G5V-IV/K0IV. 
It is in synchronous rotation with a photometric period
of 11.6626 days (Howell et al. 1986).
In Table~\ref{tab:par} we show the adopted
stellar parameters from 
Strassmeier et al. (1993) except for T$_{\rm conj}$ (see below).
 
\subsection{Radial velocities}

In our high resolution echelle spectrum we can resolve 
both components in all the orders.
We have calculated the radial velocity difference for this observation
using 39 lines clearly seen in the echelle spectrum as double lines.
The value obtained is 16.17$\pm$0.72 km s$^{-1}$ which results in 
a value for the orbital phase of 0.944$\pm$0.002.

According to the ephemeris  
given by Strassmeier et al. (1993) the calculated phase for 
this spectrum is 0.997, which is clearly in disagreement 
with our observation, since 
at this phase the radial velocity difference is 0.92  km s$^{-1}$ 
which yields a difference in wavelength between both components 
of only 0.02~\AA.
This $\Delta$$\lambda$ is lower than the resolution of the spectrum and 
then the lines from both components should be blended.
Similar problems have been found in the other spectra when we use this
ephemeris. The phase deviation is the same for all the observations and is 
very large to be ascribed to orbital period variations, so we thought of 
an error in the heliocentric Julian date on conjunction (T$_{\rm conj}$). 
In effect, we have found a mistake in the T$_{\rm conj}$ 
given in the Strassmeier et al. (1993) catalog 
where the orbital data by Griffin (1985) 
are compiled. In the original paper the epoch is determined using Modified 
Julian Date (MJD) and so a difference of 0.5 days must be added to the
Strassmeier et al. (1993) date which used 
2440000.0 Julian day as reference date.
The orbital phase of the echelle spectrum obtained using 
the corrected value of T$_{\rm conj}$ (2445737.165) is 0.954.
Now, the difference with the orbital phase that we deduce from our 
spectrum is only 0.01, which is within the uncertainties given by
Griffin (1985).
We would notice that the large discrepancies in the radial velocities 
which Gunn et al. (1996) found could be reduced using the corrected  
T$_{\rm conj}$. 
Therefore we have used the corrected T$_{\rm conj}$ 
given in Table~\ref{tab:par} 
to calculate the orbital phases of all the observations reported 
in this paper.


\begin{figure*}
{\psfig{figure=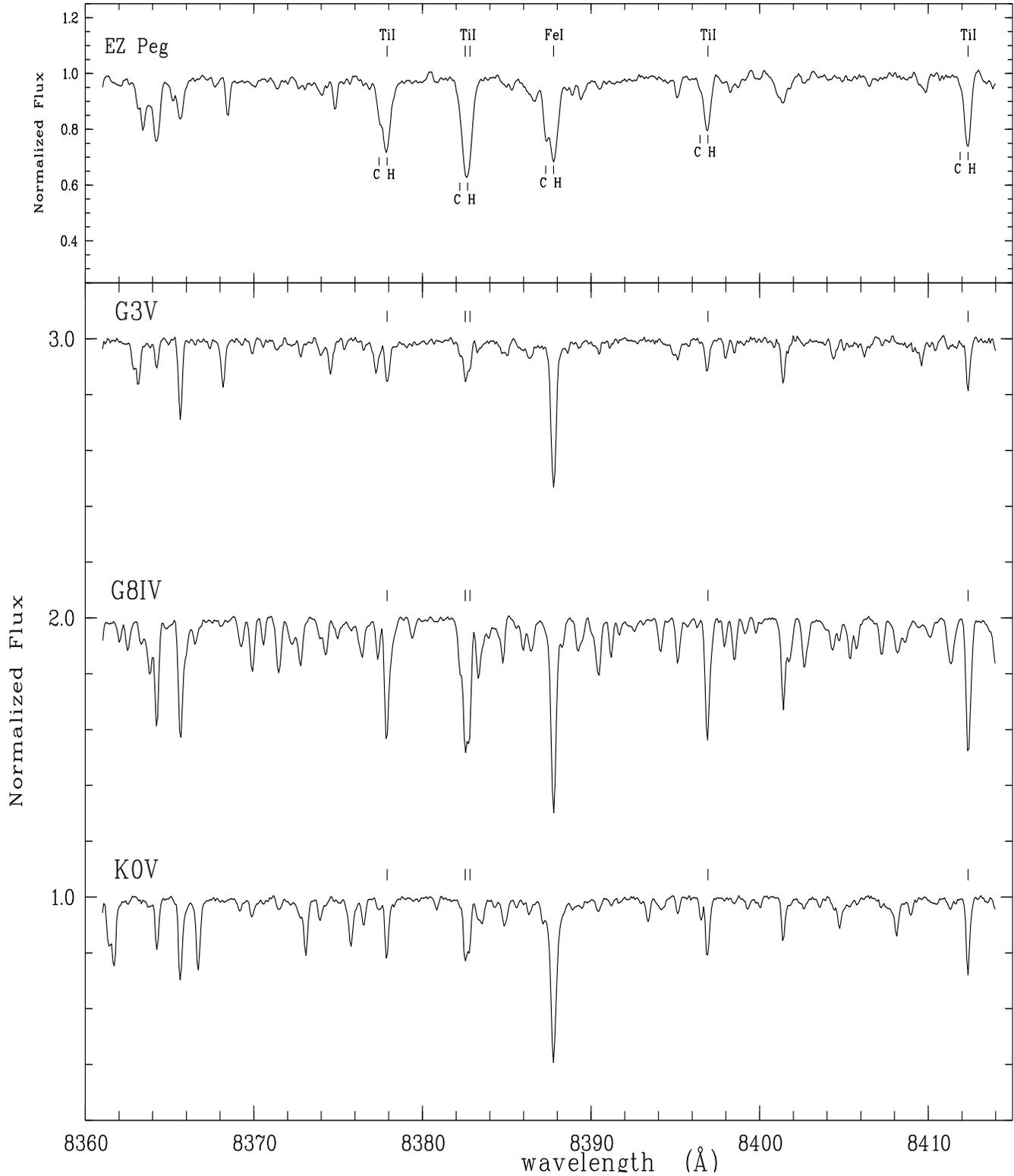,bbllx=16pt,bblly=162pt,bburx=585pt,bbury=684pt,height=21.0cm,width=18.0cm,clip=}}
\caption[ ]{WHT/UES (Jul 1993), spectra in the region of several lines of
Ti I (M.33). In the upper panel we plot EZ Peg and in the lower panel three
reference stars: HD 186427 (G3V), HD 44867 (G8IV), HD 185144 (K0V).
\label{fig:ues93ti} }
\end{figure*}

\subsection{Spectral type and luminosity class}


Several Ti~{\sc i} lines can be used as a 
luminosity classification criterium
(Keenan \& Hynek 1945; Kirkpatrick et al. 1991; 
Ginestet et al. 1994;  Jashek \& Jashek 1995) because they present 
a positive luminosity effect (i.e. they are stronger in giants than in dwarfs). 
We have analysed the behaviour of the Ti~{\sc i} lines that 
appear in our echelle spectrum in order to improve the 
luminosity classification of the EZ Peg components. 

In Fig.~\ref{fig:ues93ti} we can notice the enhancement of 
the multiplet 33 Ti~{\sc i} lines for the system
in comparison with main sequence stars 
of spectral type similar to the components of EZ Peg.
It is worth mentioning the case of the Ti~{\sc i} $\lambda$8382 
line which is even more intense than the Fe~{\sc i} $\lambda$8388 line,
showing a different behaviour that in the subgiant reference
star (see Fig.~\ref{fig:ues93ti} lower panel).
This behaviour can be also observed in other Ti~{\sc i} lines like 
$\lambda$5366.651 (Multiplet 35), $\lambda$5446.593 (M. 3), 
$\lambda$5866.453, $\lambda$5937.806 and $\lambda$5941.755 (M. 72),
$\lambda$6554.226 and $\lambda$6556.066 (M.102), $\lambda$6599.112
(M.19).
These results suggest a clear difference in luminosity class between the
two components of the system, being the G5 primary star more evolved than a 
subgiant, meanwhile the K0 star is probably a luminosity class V star.

As is explained in the following section all the activity indicators 
analysed here show that the
hot component is the active star of the system, 
contrary to the usual behaviour observed in chromospherically active binaries.
Only in some BY Dra stars the hot
component tends to be the most active star of the system 
or even the only active component (Montes et al. 1996a). 
So we have used our Ca~{\sc ii} K spectra to determine 
the absolute visual magnitude, M$_{\rm V}$ of the active component
by the application of the Wilson-Bappu effect (Wilson \& Bappu 1957).
The mean emission line width measured in our  Ca~{\sc ii} K spectra is
1.229~\AA, which come out to 60.39 km s$^{-1}$ after the quadratic 
correction of the instrumental profile.
We have used the relation found for chromospherically active binaries
by Montes et al. (1994) obtaining M$_{\rm V}$ = 2.27.
This value is lower than the  M$_{\rm V}$ 
that corresponds to a G5V
(M$_{\rm V}$(T$_{\rm sp}$) = 5.1 from
Landolt-B\"{o}rnstein (Schmidt-Kaler 1982)) but higher than for a G5III
(M$_{\rm V}$(T$_{\rm sp}$) = 0.9).
This result indicates that the active component of EZ Peg may  
be of luminosity class IV or higher.

The difference in luminosity class 
between the hot and cool components
that our spectra indicate
could explain why the hot component is the active star (see below),
since if the hot component is a more evolved star it could have 
developed a deep convective zone and therefore, 
according to the dynamo mechanism,
should present a higher activity level.
Anyway the cooler K0V component could also be active 
and present a faint emission in the H \& K lines, but due to the small
contribution of that star to the combined spectrum of the system, this 
emission is very difficult to be observed.  

\subsection{The Li~{\sc i} $\lambda$6707.8 line}

\begin{figure*}
{\psfig{figure=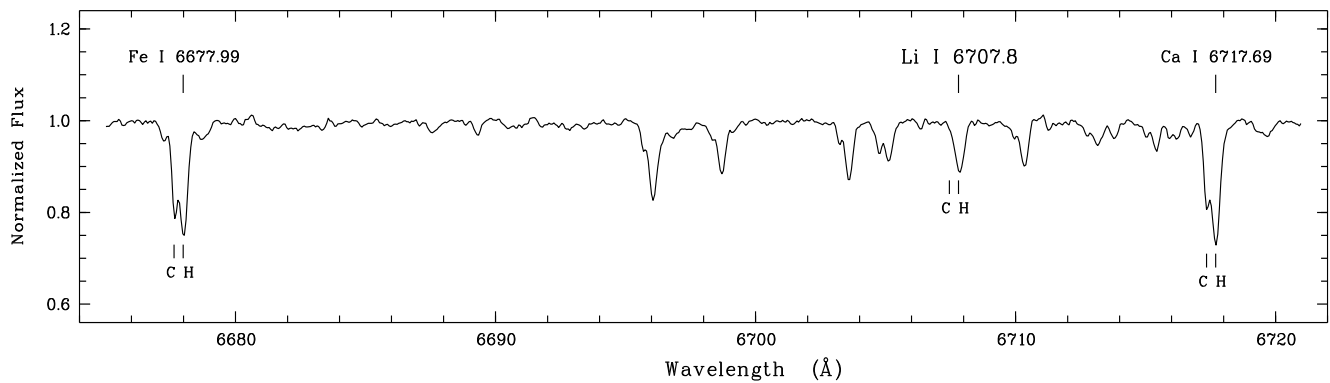,bbllx=28pt,bblly=284pt,bburx=585pt,bbury=448pt,height=5.6cm,width=18.0cm,clip=}}
\caption[ ]{WHT/UES (Jul 1993), spectrum in the region of the Li I 6707.8~\AA$\ $ line
including also the Fe I 6677.99~\AA$\ $ and Ca I 6717.69~\AA$\ $. The
expected positions of this features for the hot (H) and cool (C) components
are given.
\label{fig:ues93li} }
\end{figure*}

We also analyse here the Lithium abundance of EZ Peg using the 
Li~{\sc i} $\lambda$6707.8 line.
This line appears in the order 33 of the WHT/UES echelle spectrum
together with the Fe~{\sc i} $\lambda$6678 and Ca~{\sc i} $\lambda$6718 lines.  
In Fig.~\ref{fig:ues93li} we can clearly see the  Fe~{\sc i} and  Ca~{\sc i}
lines from both components, the expected positions of these features for 
the hot (H) and cool (C) are given in this figure.
However, the  absorption Li~{\sc i} line appears centered in the 
wavelength-position corresponding to the hot component, so
we can suppose that the absorption line comes only from this star or that the 
contribution of the cool component to the observed Li~{\sc i} is very small.
This assumption seems to be reasonable because, as we show below, the hot 
component is the active star of the system and  
a high Li~{\sc i} abundance is found in 
chromospherically active binaries (Fern\'{a}ndez-Figueroa et al. 1993;
Barrado et al. 1997).
In order to obtain the Li~{\sc i} equivalent width (EW), we have corrected 
the total EW measured, EW(Li~{\sc i}+Fe~{\sc i})=50.4 (m\AA)
from the relative contribution to the continuum (0.65). Then  
the contribution
from the blend with the Fe~{\sc i} $\lambda$ 6704.41 line must be subtracted.
The EW of the Fe~{\sc i} line was calculated from the empirical 
relationship with (B-V) given by Soderblom et al. (1990), obtaining
EW(Fe~{\sc i})=15.8 (m\AA), the same value is found with the relationship 
given by Favata et al. (1993) using the effective temperature.
Finally, the corrected  EW(Li~{\sc i})$_{\rm corr}$=61.7 (m\AA) was converted
into abundances by means of the curves of growth computed by Pallavicini et al. 
(1987), obtaining log~N(Li~{\sc i})=2.15 (on a scale where log~N(H)=12.0) 
with an accuracy of $\approx$~0.30~dex.
This high Li~{\sc i} abundance is similar to that obtained for other
chromospherically active binaries  by Fern\'{a}ndez-Figueroa et al. (1993) 
and Barrado et al. (1997) and it is related to the high activity level of
EZ Peg.


\section{Chromospheric activity indicators}

In the following we describe the behaviour of the different optical
chromospheric activity indicators (formed at different atmospheric heights) 
observed for EZ Peg: 
Na~{\sc i} D$_{1}$, D$_{2}$, and Mg~{\sc i} b triplet
(upper photosphere and lower chromosphere),
Ca~{\sc ii} IRT lines (lower chromosphere), 
H$\alpha$, H$\beta$, Ca~{\sc ii} H \& K (middle chromosphere), and
He~{\sc i} D$_{3}$ (upper chromosphere).
The chromospheric contribution in
these features has been determined  using
the spectral subtraction technique described in
detail by Montes et al. (1995a, b, c), (see also Paper I).
The synthesized spectrum has been constructed with reference stars of
spectral types G5V and K0IV, taken from the
spectral library of Montes et al. (1997a), in the case of the INT/IDS spectra.
For the WHT/UES spectrum we have used a G3V and K0V taken during the same 
observing run, and for the JKT/RBS spectrum a G5III-IV and K0V 
have been used.
The use of reference spectra of spectral type slightly different for 
each observing run and different also from the system component spectra 
introduce some
residual in the absorption lines, but it does not affect 
significatively the excess 
emission equivalent width measures in the subtracted spectra.
The uncertainties are within similar order to that obtained in the reduction,
 synthesis and fitting procedures. Furthermore, we have considered a 
clear detection of excess emission or absorption lines only when these 
features in the difference spectrum are larger than 3 $\sigma$.
 
The contribution of the hot and cool components
to the total continuum that yields a better fit between the 
observed and subtracted spectra is 0.65/0.35.

The line profiles are displayed in Figs.~1~-~8.
For each spectrum we plot the observed (solid-line), the
synthesized (dashed-line) and the subtracted one, additively
offset for better display (dotted line).

\begin{table*}
\caption[]{H$\alpha$ line measures in the observed and
subtracted spectrum
\label{tab:measuresha}}
\begin{flushleft}
\scriptsize
\begin{tabular}{lcccccccccccc}
\noalign{\smallskip}
\hline
\noalign{\smallskip}
 &   &  
\multicolumn{3}{c}{Observed H$\alpha$ Spectrum} &\ &
\multicolumn{4}{c}{Subtracted H$\alpha$ Spectrum} \\
\cline{3-5}\cline{7-10}
\noalign{\smallskip}
 {Obs.} &  {$\varphi$} &  
 {W$_{\rm obs}$} & {R$_{\rm c}$ } & {F(1.7\AA)} &  &
 {W$_{\rm sub}$} & {I} & {EW} & {$\log {\rm F}_{\rm S}$} \\
  &    & {\scriptsize (\AA) } &  &  & &
 {\scriptsize (\AA)} &  & {\scriptsize  (\AA)} &  \\
\noalign{\smallskip}
\hline
\noalign{\smallskip}
%
UES 1993 & 0.954 & 0.96 & 1.325 & 1.987 & &
 1.62 & 0.882 & 2.085 & 7.06 \\
\noalign{\smallskip}
INT 1995 & 0.416 & 0.65 & 1.139 & 1.695 & &
 1.70 & 0.655 & 1.488 & 6.91 \\
   "     & 0.501 & 0.54 & 1.059 & 1.657 & &
 1.69 & 0.639 & 1.400 & 6.89 \\
\noalign{\smallskip}
INT 1996 & 0.141 & 1.48 & 1.131 & 1.779 & &
 2.43  & 0.604 & 2.272 & 7.10 \\
   "     & 0.224 & 1.55 & 1.157 & 1.812 & &
 2.30  & 0.634 & 1.862 & 7.01 \\
   "     & 0.306 & 1.35 & 1.134 & 1.784 & &
 2.16  & 0.638 & 1.677 & 6.96 \\
%
\noalign{\smallskip}
\hline
\noalign{\smallskip}
\end{tabular}
\end{flushleft}
\end{table*}

\begin{table*}
\caption[]{Parameters of the broad and narrow Gaussian components
used in the fit of the H$\alpha$ subtracted spectra 
\label{tab:measuresha_nb}}
\begin{flushleft}
\scriptsize
\begin{tabular}{lcccccccccc}
\noalign{\smallskip}
\hline
\noalign{\smallskip}
\noalign{\smallskip}
        &    &  
\multicolumn{4}{c}{H$\alpha$ broad component} &\ &
\multicolumn{4}{c}{H$\alpha$ narrow component} \\
\cline{3-6}\cline{8-11}
\noalign{\smallskip}
 {Obs.} & {$\varphi$} & 
{I} & FWHM & EW$_{\rm B}$ & EW$_{\rm B}$/EW$_{\rm T}$ & &  
{I} & FWHM & EW$_{\rm N}$ & EW$_{\rm N}$/EW$_{\rm T}$ \\
        &           & 
    & {\scriptsize (\AA)} & {\scriptsize (\AA)} & (\%) & & 
    & {\scriptsize (\AA)} & {\scriptsize (\AA)} & (\%) \\
\noalign{\smallskip}
\hline
\noalign{\smallskip}
%
%
UES 1993 & 0.954 & 0.273 & 3.935 & 1.143 & 54.8 & & 0.648 & 1.367 & 0.942 & 45.2 \\
\noalign{\smallskip}
INT 1995 & 0.416 & 0.152 & 4.035 & 0.651 & 43.7 & & 0.535 & 1.470 & 0.837 & 56.3 \\
  "      & 0.501 & 0.129 & 4.096 & 0.562 & 40.1 & & 0.535 & 1.470 & 0.838 & 59.9 \\ 
\noalign{\smallskip}
INT 1996 & 0.141 & 0.123 & 9.418 & 1.209 & 53.2 & & 0.493 & 2.028 & 1.064 & 46.8 \\
  "      & 0.224 & 0.124 & 5.622 & 0.740 & 39.7 & & 0.520 & 2.028 & 1.122 & 60.3\\
  "      & 0.306 & 0.091 & 5.528 & 0.536 & 32.0 & & 0.572 & 1.876 & 1.142 & 68.0\\ 
\noalign{\smallskip}
\hline
\noalign{\smallskip}
\end{tabular}

\end{flushleft}
\end{table*}


\begin{figure*}
{\psfig{figure=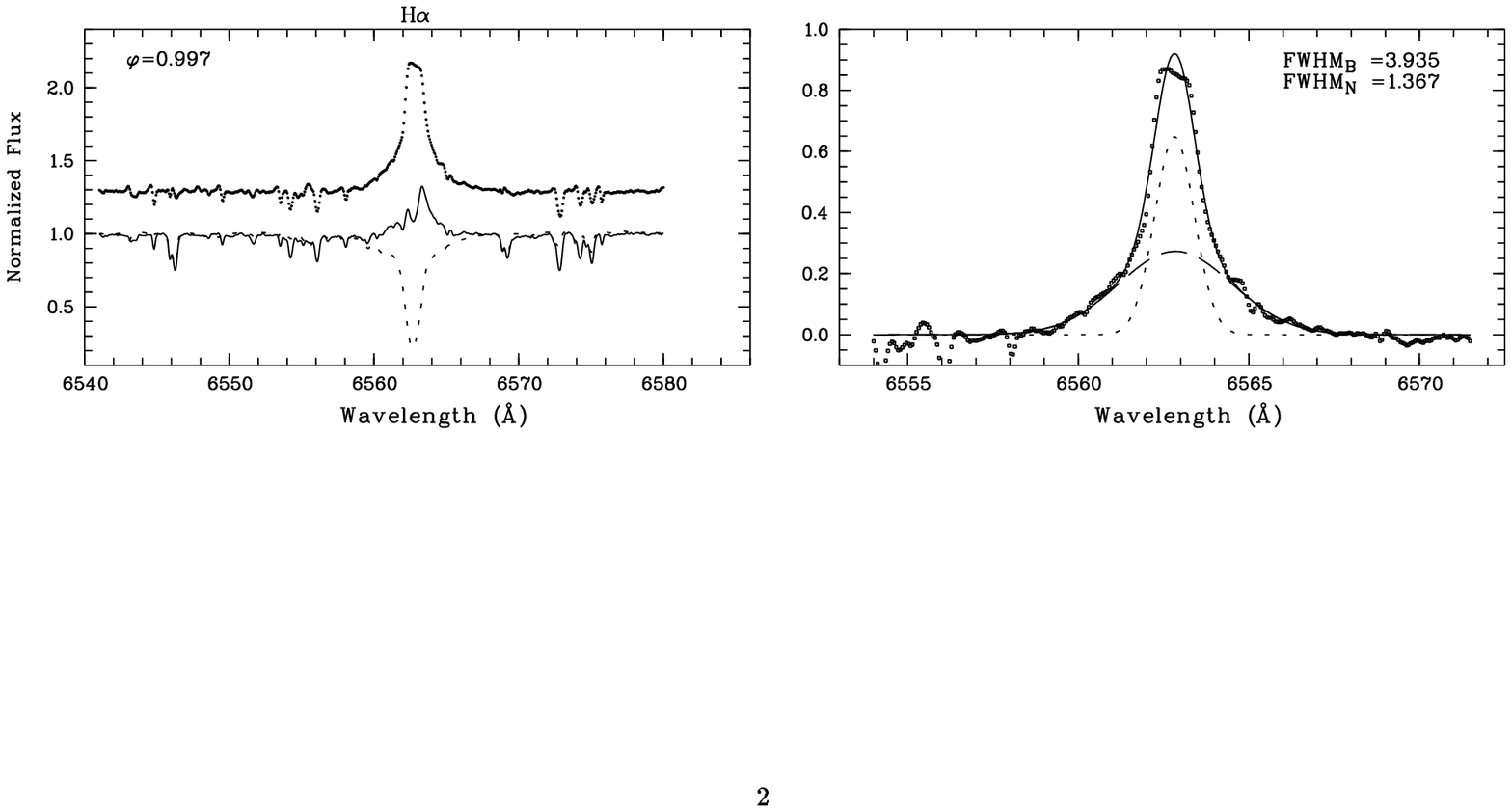,bbllx=28pt,bblly=284pt,bburx=578pt,bbury=448pt,height=5.6cm,width=18.0cm,clip=}}
\caption[ ]{WHT/UES H$\alpha$ spectrum (Jul 1993).
In the left panel  we plot the observed spectrum (solid-line), the
synthesized spectrum (dashed-line), the subtracted spectrum, additively
offset for better display (dotted line).
In the right panel we plot the subtracted H$\alpha$ profile (dotted line).
and the the two Gaussian components fit (solid-line).
The sort-dashed-line represents the broad component
and the large-dashed-line the narrow one.
\label{fig:ues93ha} }
\end{figure*}

\begin{figure*}
{\psfig{figure=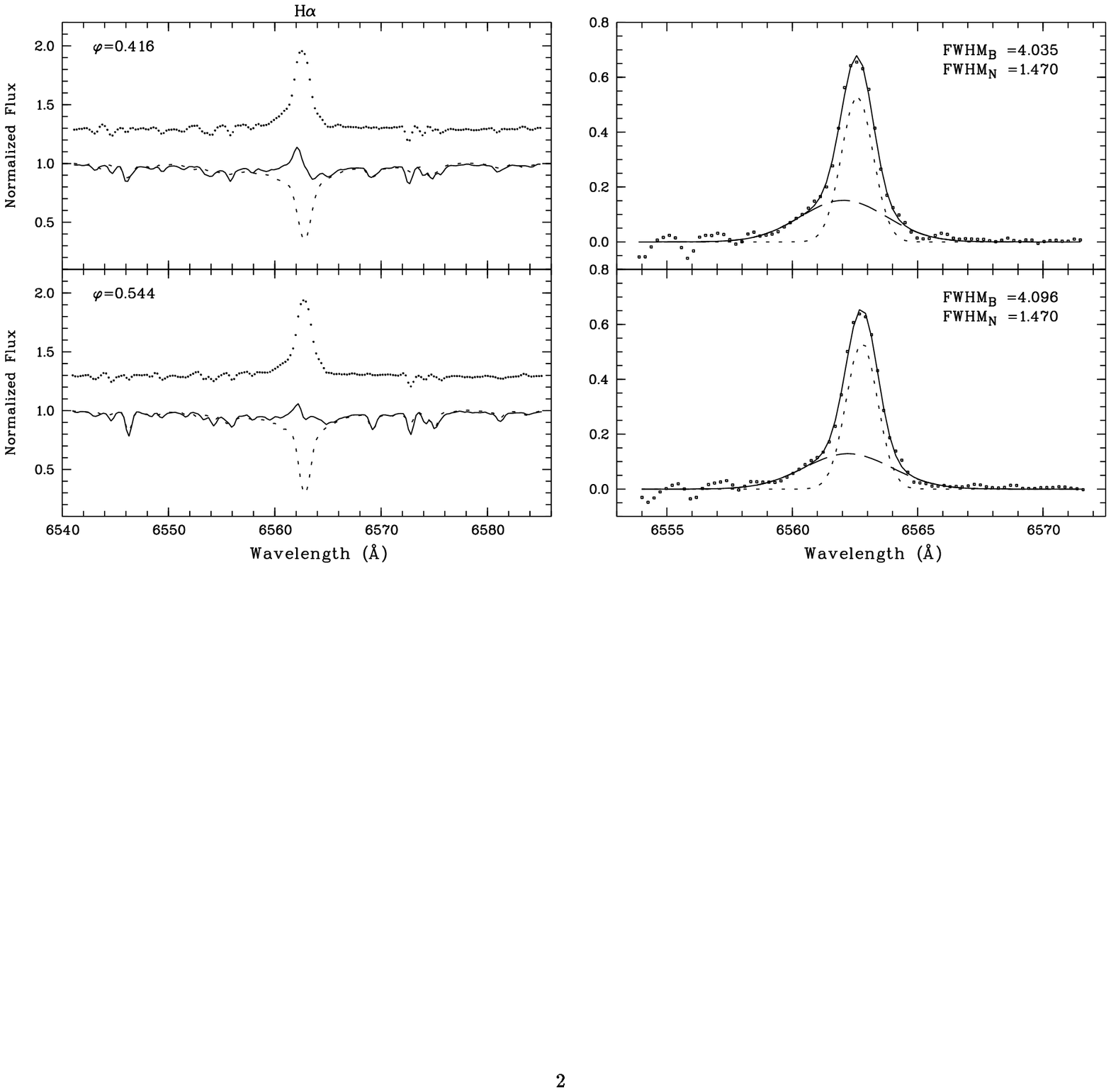,bbllx=28pt,bblly=404pt,bburx=578pt,bbury=688pt,height=10.0cm,width=18.0cm,clip=}}
\caption[ ]{INT/IDS H$\alpha$ spectra (Sep 1995),
as in Fig~.\ref{fig:ues93ha}
\label{fig:int95ha} }
\end{figure*}

\begin{figure*}
{\psfig{figure=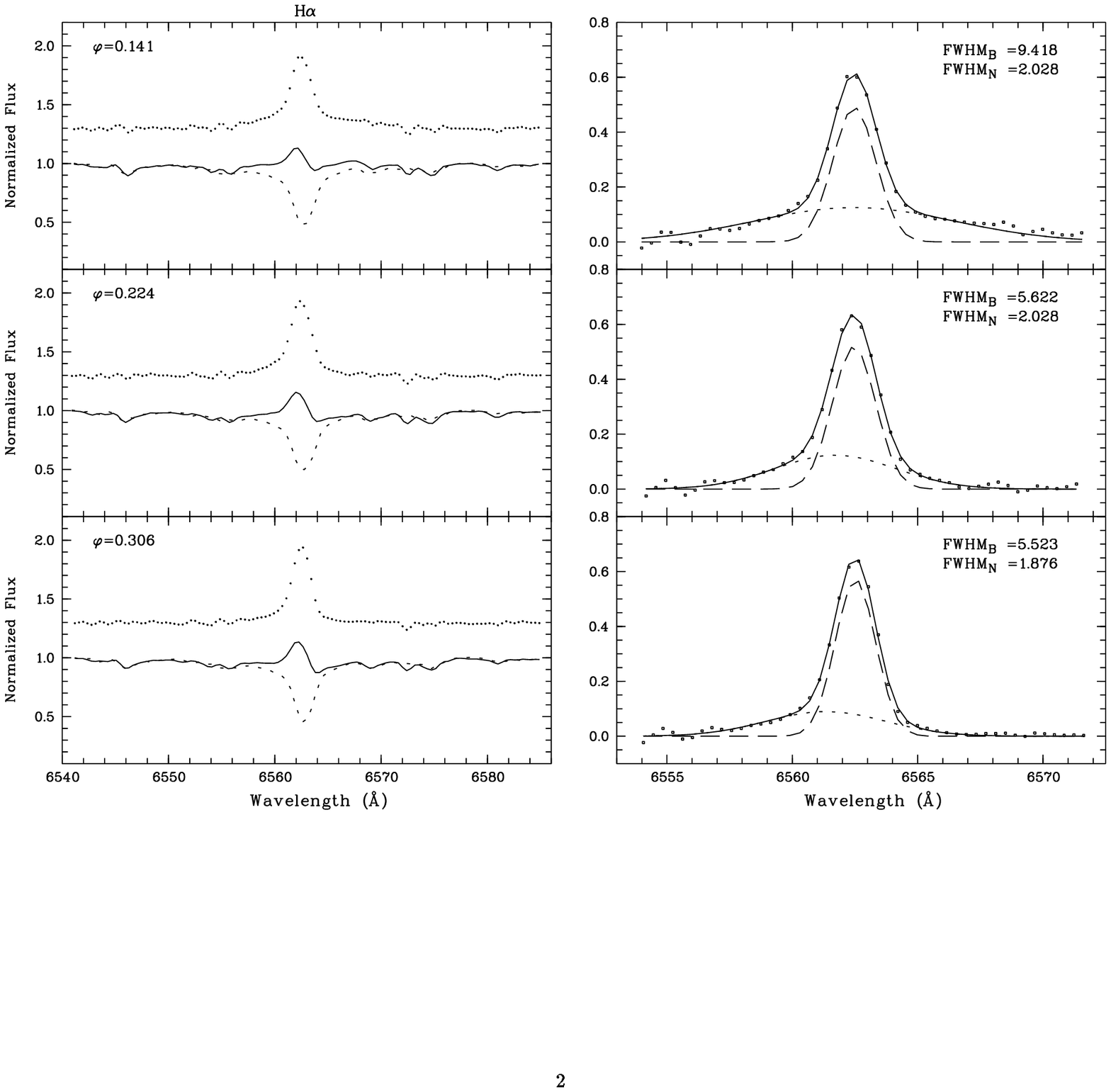,bbllx=28pt,bblly=284pt,bburx=578pt,bbury=688pt,height=15.0cm,width=18.0cm,clip=}}
\caption[ ]{INT/IDS H$\alpha$ spectra (Nov 1996),
as in Fig~.\ref{fig:ues93ha}
\label{fig:int96ha} }
\end{figure*}

\subsection{The H$\alpha$ line}

We have taken several spectra of EZ Peg in the H$\alpha$ line region
in three different epochs. One WHT/UES spectrum in July 1993 at the 
orbital phase 0.954 (Fig.~\ref{fig:ues93ha}), 
two INT/IDS spectra in September 1995 at the
phases 0.416 and 0.501 (Fig.~\ref{fig:int95ha})
and three more INT/IDS spectra in November 1996 
at the phases 0.141, 0.224 and 0.306 (Fig.~\ref{fig:int96ha}).
In all the spectra we can see the H$\alpha$ line in emission above the 
continuum with an asymmetry in the wavelength position corresponding to the
H$\alpha$ absorption line of the cool component.
The spectral subtraction reveals that the hot star is responsible for 
the excess H$\alpha$ emission.

Table~\ref{tab:measuresha} 
gives the  H$\alpha$ line parameters, measured in the
observed and subtracted spectra of EZ Peg as follows: 
column (1) the observing run,
column (2) the orbital phase ($\varphi$) and
columns (3), (4), (5) the following parameters
measured in the observed spectrum:
the full width at half maximum (W$_{\rm obs}$);
the residual intensity, R$_{{\rm c}}$; and the H$\alpha$ core flux, F(1.7\AA),
measured as the residual area below the central 1.7~\AA$\ $ passband.
The last four columns give the following parameters
measured in the subtracted spectrum:
the full width at half maximum
(W$_{\rm sub}$), the peak emission intensity (I),
 the excess H$\alpha$ emission equivalent width (EW( H$\alpha$)), and
absolute fluxes at the stellar surface logF$_{\rm S}$(H$\alpha$) obtained
with the calibration of Pasquini \& Pallavicini (1991) as a function of
(V~-~R), very similar values of F$_{\rm S}$(H$\alpha$) are obtained
using the more recent calibration of Hall (1996)  as a function of
(V~-~R) and (B~-~V).
As can be seen in Table~\ref{tab:measuresha} and 
Figures \ref{fig:ues93ha}, \ref{fig:int95ha} and \ref{fig:int96ha} the 
excess H$\alpha$ emission of EZ Peg  show small variations with the orbital 
phase and also seasonal variations from Jul 93 to Nov 96.
The higher H$\alpha$ EW has been reached in the first night of the 
1996 run and in the 1993 observation.
The higher values correspond to the orbital phases closer to the 
conjunction (0.0). 

The H$\alpha$ subtracted profiles present 
at all the epochs broad wings which
are not well matched using a single-Gaussian fit.
These profiles have therefore been fitted using two Gaussian components.
The parameters of the broad and narrow components used in the two-Gaussian fit
are given in table~\ref{tab:measuresha_nb} and the corresponding profiles are
plotted in the left panel of 
Figures \ref{fig:ues93ha}, \ref{fig:int95ha} and \ref{fig:int96ha}.
In some cases the blue wing is noticeable stronger than the red wing and the
fit is better matched when the broad component is blue-shifted with respect 
to the narrow component. The cool component of EZ Peg cannot account for 
this effect since any emission from that star would be on the red side of the
observed line for these orbital phases.
We have also observed broad components in the most active systems of 
the sample of Paper I which we interpreted as microflaring that occurs in
the chromosphere by similarity with the 
broad components also found in the chromospheric Mg~{\sc ii} h \& k lines
(Wood et al. 1996) and in several transition region lines of active stars
(Linsky \& Wood 1994; Linsky et al. 1995;
Wood et al. 1996, 1997; Dempsey et al. 1996a, b; Robinson et al. 1996).
The contribution of the broad component to the total EW of the line 
of EZ Peg ranges between 32\% and 55\% which
is in the range observed to the stars analysed in Paper I.
The larger changes in the excess
H$\alpha$ emission appear to occur
predominantly in the broad component and 
its contribution is related to the degree of stellar activity.
It is noticeable the change in the wings of the H$\alpha$ line from the 
1st to the 2nd night of the 1996 run, the 
two Gaussian components fit reveals that the contribution of 
the broad component changes from 53.2\% to 39.7\%
(see Fig~\ref{fig:int96ha}).


\begin{figure*}
{\psfig{figure=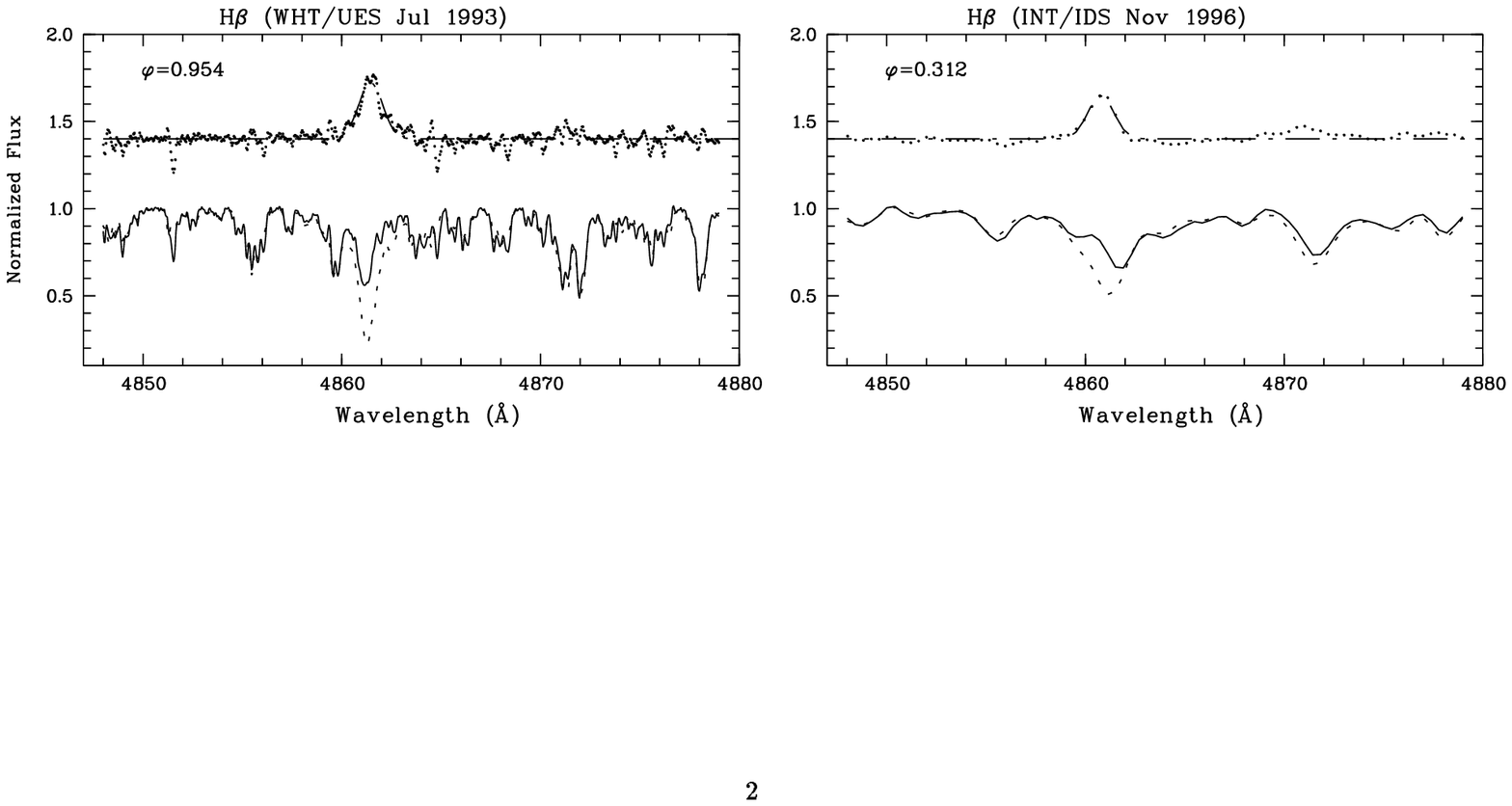,bbllx=28pt,bblly=284pt,bburx=585pt,bbury=448pt,height=5.6cm,width=18.0cm,clip=}}
\caption[ ]{WHT/UES H$\beta$ (Jul 1993) and INT/IDS H$\beta$  (Nov 1996)
spectra.
For each spectrum we plot the observed spectrum (solid-line), the
synthesized spectrum (dashed-line), the subtracted spectrum, additively
offset for better display (dotted line).
and the Gaussian fit to the subtracted spectrum (dotted-dashed line)
\label{fig:ues93hb} }
\end{figure*}

\subsection{The H$\beta$ line}

Two spectra in the H$\beta$ line region are available.
The WHT/UES spectrum (July 1993) and other in the INT/IDS 1996 run.
In both spectra the H$\beta$ line appears in absorption and the 
application of the spectral subtraction technique reveals
a clear excess emission in this line (see Fig.~\ref{fig:ues93hb}).
Again the relative velocity shift shows that the emissions belong to the 
hotter component. Contrary to the case of the H$\alpha$ line 
the H$\beta$ subtracted profiles do not present 
broad wings and they are well matched using a single-Gaussian fit.

Table~\ref{tab:measureshb} 
gives the  H$\beta$ line parameters, measured in the
observed and subtracted spectra as in the case of the H$\alpha$ line, 
see also our previous study of the excess H$\beta$ emission 
in active binaries (Montes et al. 1995d).

\begin{table*}
\caption[]{H$\beta$ line measurements in the observed and subtracted spectrum
\label{tab:measureshb} }
\begin{flushleft}
\scriptsize
\begin{tabular}{ l c c c c c c c c c c c c c c}
\noalign{\smallskip}
\hline
\noalign{\smallskip}
   &    &    \multicolumn{3}{c}{Observed H$\beta$ Spectrum} &\ &
\multicolumn{3}{c}{Subtracted H$\beta$ Spectrum} & & \\
\cline{3-5}\cline{7-9}
\noalign{\smallskip}
{Obs.} & {$\varphi$} &
 { W$_{\rm obs}$ } & { R$_{\rm c}$ } & { F(1.7\AA) } &  &
 { W$_{\rm sub}$ } & { I} & { EW } &
$\frac{\rm EW(H\alpha)}{\rm EW(H\beta)}$ &
$\frac{\rm E_{H\alpha}}{\rm E_{H\beta}}$ \\
  &     &  {\scriptsize (\AA) } &  &  & &
 {\scriptsize (\AA)} &  & {\scriptsize  (\AA)} &  &  \\
\noalign{\smallskip}
\hline
\noalign{\smallskip}
 &       &  \\
UES 1993   & 0.954 & 0.91 & 0.561 & 1.235 &  & 
0.94 & 0.364 & 0.489 & 4.27 & 3.49 \\
\noalign{\smallskip}
INT 1996   & 0.312 & 2.13 & 0.660 & 1.237 &  & 
1.54 & 0.244 & 0.411 & 4.08 & 3.33 \\
%
   &       &  \\
\noalign{\smallskip}
\hline
\noalign{\smallskip}
\end{tabular}
\end{flushleft}
\end{table*}

In Table~\ref{tab:measureshb} we also give the ratio of excess emission EW,
$\frac{\rm EW(H\alpha)}{\rm EW(H\beta)}$, for the two H$\beta$ observations
of EZ Peg, and the ratio
of excess emission $\frac{\rm E_{H\alpha}}{\rm E_{H\beta}}$
with the correction:

\[ \frac{\rm E_{H\alpha}}{\rm E_{H\beta}} =
\frac{\rm EW(H\alpha)}{\rm EW(H\beta)}*0.2444*2.512^{(B-R)}.\]
given by Hall \& Ramsey (1992) that takes into account the absolute flux
density in these lines and the color difference in the components.
                  
We have used this ratio as a diagnostic for
discriminating between the presence of plages and prominences in the stellar
surface, following the results of Hall \& Ramsey (1992) 
who found that low E$_{\rm H\alpha}$/E$_{\rm H\beta}$  ($\approx$~1-2)
can be achieved both in plages and prominences viewed against the disk, but
that high ratios ($\approx$~3-15) can only be achieved in extended regions
viewed off the limb.
The high ratio (E$_{\rm H\alpha}$/E$_{\rm H\beta}$  $>$3, 
see Table~\ref{tab:measureshb}) 
that we have found in EZ Peg indicates 
that the emission would arise from extended regions (prominences).

The excess H$\beta$ emission is somewhat larger in the 1993 observation
than in the last night observation of the 1996 run, in agreement with the 
behaviour observed for the H$\alpha$ line.

\subsection{The Na~{\sc i} D$_{1}$ and D$_{2}$ and He~{\sc i} D$_{3}$ lines}

\begin{figure*}
{\psfig{figure=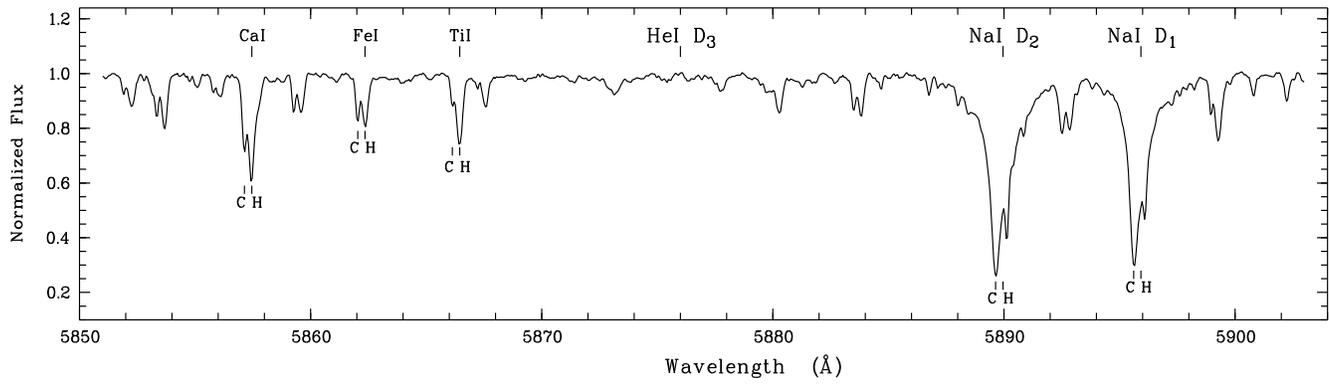,bbllx=28pt,bblly=284pt,bburx=585pt,bbury=448pt,height=5.6cm,width=18.0cm,clip=}}
\caption[ ]{WHT/UES (Jul 1993), spectrum in the region of the NaI and HeI lines
including also the CaI 5857.454~\AA, FeI 5862.357~\AA$\ $ and TiI 5866.453\AA.
 The expected positions of these features for the hot (H) and
 cool (C) components are given.
\label{fig:ues93he} }
\end{figure*}

\begin{figure*}
{\psfig{figure=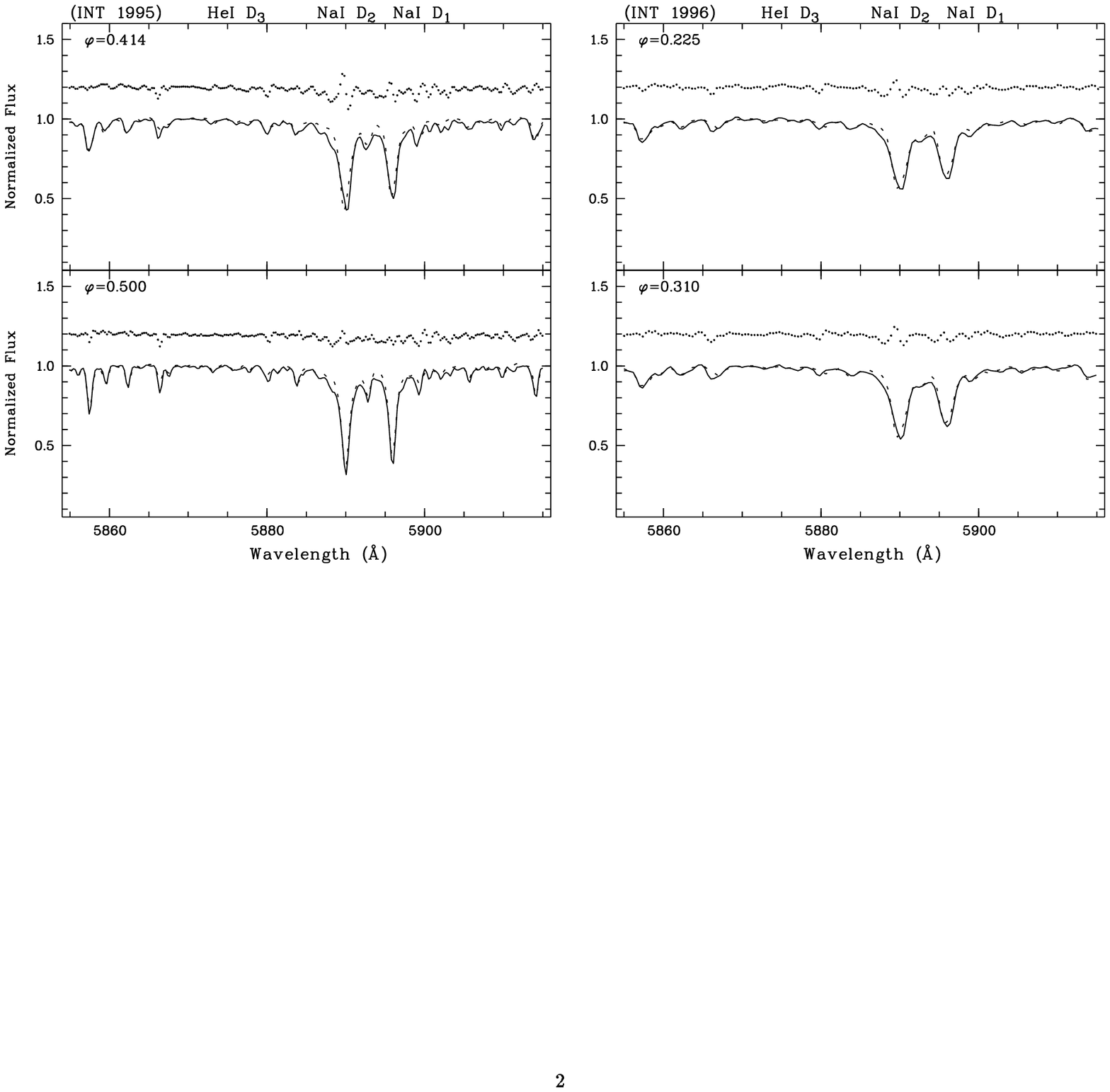,bbllx=28pt,bblly=404pt,bburx=578pt,bbury=688pt,height=10.0cm,width=18.0cm,clip=}}
\caption[ ]{INT/IDS spectra in the region of the NaI and HeI lines
(Sep 1995) and (Nov 1996).
For each spectrum we plot the observed spectrum (solid-line), the
synthesized spectrum (dashed-line), the subtracted spectrum, additively
offset for better display (dotted line).
\label{fig:int9596he} }
\end{figure*}

We have five spectra of EZ Peg in the 
region of  Na~{\sc i} D$_{1}$ and D$_{2}$ and He~{\sc i} D$_{3}$ lines: 
one WHT/UES spectrum in July 1993 at the orbital phase 0.954 
(Fig.~\ref{fig:ues93he}), 
two INT/IDS spectra in September 1995 at the phases 0.414 and 0.500
and two spectra in INT/IDS November 1996 at the phases 0.225 and 0.310
(Fig.~\ref{fig:int9596he}). 
We have not detected
any significant absorption or emission for He~{\sc i} D$_{3}$. 

Saar et al. (1997) in their study of He~{\sc i} D$_{3}$ in G and K
dwarfs found that while for few active stars the He~{\sc i} D$_{3}$  line
behaves "normally", increasing in absorption with increasing rotation, 
and showing consistent correlations with other activity indicators, 
the behavior clearly diverges when stars become very active.
They found a large filling-in He~{\sc i} D$_{3}$ line 
in some very active K stars.
On the other hand, they also found He~{\sc i} D$_{3}$ EW significantly lower
than the theoretical maximum value given by Andretta \& Giampapa (1995), 
suggesting that the line could be filling-in due to 
frequent low-level flaring.

Both facts could explain the absence of emission or absorption due to 
a total filling-in of the He~{\sc i} D$_{3}$ line that we have found
in EZ Peg and also in other dwarfs and subgiants 
chromospherically active binaries (see Paper I), since these stars are very 
active and also present low-level flaring (microflares) as indicated by the
H$\alpha$ broad component that we have found (see sect. 4.1 and Paper I).
In some cases He~{\sc i} D$_{3}$ goes into emission due to flares
(Huenemoerder \& Ramsey 1987; Huenemoerder et al. 1990; 
Montes et al. 1996b; Paper I). However,
in the most evolved stars the behaviour is different as consequence of 
the lower chromospheric densities of these stars, as it is suggested 
by Saar et al. (1997)
and as it is indicated by the presence of 
the He~{\sc i} D$_{3}$ in absorption that 
we have reported in Paper I for several chromospherically active binaries 
with giant components.


In the Na~{\sc i} D$_{1}$ and D$_{2}$ lines
the spectra are too noisy to distinguish a clear filled-in. 
Once again we can observe (see Fig.~\ref{fig:ues93he}) 
that the redward lines 
of the double-lined spectrum usually
are the deepest, but in the Na~{\sc i} D whose blueshifted components
are the deepest ones. As it is well known, the Na~{\sc i} D lines are very 
sensitive to the temperature;
this behaviour of the lines
confirms that the hot star of the system is the redshifted one,
which, as mentioned above, shows very high activity in H$\alpha$.

\subsection{The Mg~{\sc i} b triplet lines}

The Mg~{\sc i} b triplet lines $\lambda$5167.33, $\lambda$5172.68, and
$\lambda$5183.61 are formed in the lower chromosphere and the
region of temperature minimum and they are good diagnostics of
photospheric activity (Basri et al. 1989).
A small filling-in of these lines have been recently found 
in some RS CVn system (Gunn \& Doyle 1997; Gunn et al. 1997).
As in the case of the Na~{\sc i} D$_{1}$ and D$_{2}$ lines, 
in the subtracted spectrum corresponding to the echelle images 
no clear filled-in has been found. 
Thus the photospheric activity in EZ Peg is very low. 

\subsection{The Ca~{\sc ii} H $\&$ K and H$\epsilon$ lines}

\begin{figure*}
{\psfig{figure=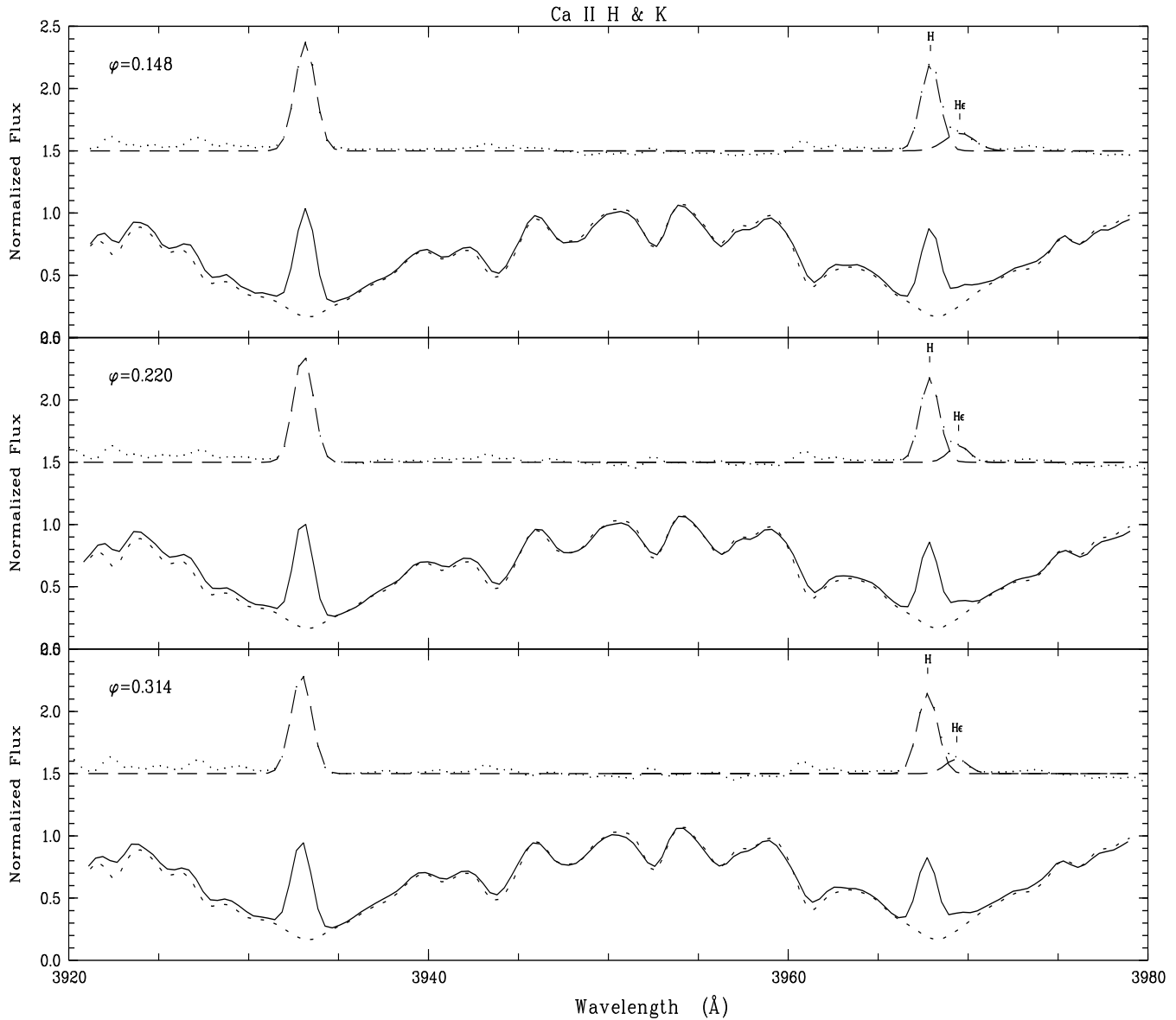,bbllx=28pt,bblly=162pt,bburx=585pt,bbury=568pt,height=16.0cm,width=18.0cm,clip=}}
\caption[ ]{INT/IDS Ca~{\sc ii} H \& K  spectra (Nov 1996),
as in Fig~.\ref{fig:ues93hb}
\label{fig:int96ca} }
\end{figure*}

We have taken three spectra in the
Ca~{\sc ii} H \& K line region during the 1996 run
(see Fig.~\ref{fig:int96ca}).
These spectra exhibit strong Ca~{\sc ii} H \& K emission lines and a small 
bump in the wavelength-position corresponding to the H$\epsilon$ line. 
The H \& K emission lines appear blue-shifted with respect to the observed 
absorption lines in agreement with the orbital phase, which confirms 
that the emission originates in the hot component.
The spectral subtraction yields a satisfactory fit and reveals a small 
but clear emission in the H$\epsilon$ line.

The excess Ca~{\sc ii} H \& K and H$\epsilon$ emissions
 equivalent widths (EW) have been measured
by reconstruction of the absorption line profile
(described by Fern\'andez-Figueroa et al. 1994) and using the
spectral subtraction technique (explained by Montes et al. 1995c, 1996a).
The Ca~{\sc ii} H \& K surface flux, F$_{\rm S}$(Ca~{\sc ii}~H \& K)
has been  obtained by means of
the linear relationship between the absolute surface
flux at 3950~\AA$\ $ (in erg cm$^{-2}$ s$^{-1}$ \AA$^{-1}$)
and the colour index (V-R) by Pasquini et al. (1988).

Table~\ref{tab:measureshyk} gives
the Ca~{\sc ii} H \& K and H$\epsilon$ line parameters, measured in the
observed and subtracted spectra.
Column (1) gives the observing run.
Column (2) gives
the orbital phase ($\varphi$).
In columns (3) and (4) we list the EW for the K and H lines,
obtained by reconstruction of the absorption line profile, 
in columns (5) to (7) we give the EW for
the K, H and H$\epsilon$ lines, measured in the subtracted
spectrum, and in columns (8) to (10) the corresponding surface fluxes
are listed.


\begin{table*}
\caption[]{Ca~{\sc ii} H \& K and H$\epsilon$ lines measures in 
the observed and subtracted spectrum  
\label{tab:measureshyk} }
\begin{flushleft}
\scriptsize
\begin{tabular}{l c c c c c c c c c c c c c}
\hline
\noalign{\smallskip}
 &     & \multicolumn{2}{c}{Reconstruction} &\ &
\multicolumn{3}{c}{Spectral subtraction} &\ &
\multicolumn{3}{c}{Absolute flux} \\
\cline{3-4}\cline{6-8}\cline{10-12}
\noalign{\smallskip}
\noalign{\smallskip}
 Obs.  & $\varphi$ &
EW & EW & &
EW & EW & EW & &
$\log{\rm F}$ & $\log{\rm F}$ & $\log{\rm F}$  \\
 & & (K) & (H)
& & (K) & (H) & (H$\epsilon$) & &
  (K) & (H) & (H$\epsilon$) \\
\noalign{\smallskip}
\hline
\noalign{\smallskip}
INT 1996  & 0.148 & 0.982 & 0.718 & & 1.246 & 0.931 & 0.292 & & 
6.91 & 6.79 & 6.28 \\
"         & 0.220 & 0.975 & 0.671 & & 1.191 & 0.910 & 0.194 & &
6.82 & 6.78 & 6.10 \\
"         & 0.314 & 0.948 & 0.628 & & 1.142 & 0.907 & 0.185 & &
6.87 & 6.77 & 6.08 \\
\noalign{\smallskip}
\hline
\end{tabular}
\end{flushleft}
\end{table*}

The excess Ca~{\sc ii} H \& K and H$\epsilon$ emissions change with the 
orbital phase during the three nights of the 1996 run 
in the same way as the corresponding excess H$\alpha$ 
emissions.


\begin{figure*}
{\psfig{figure=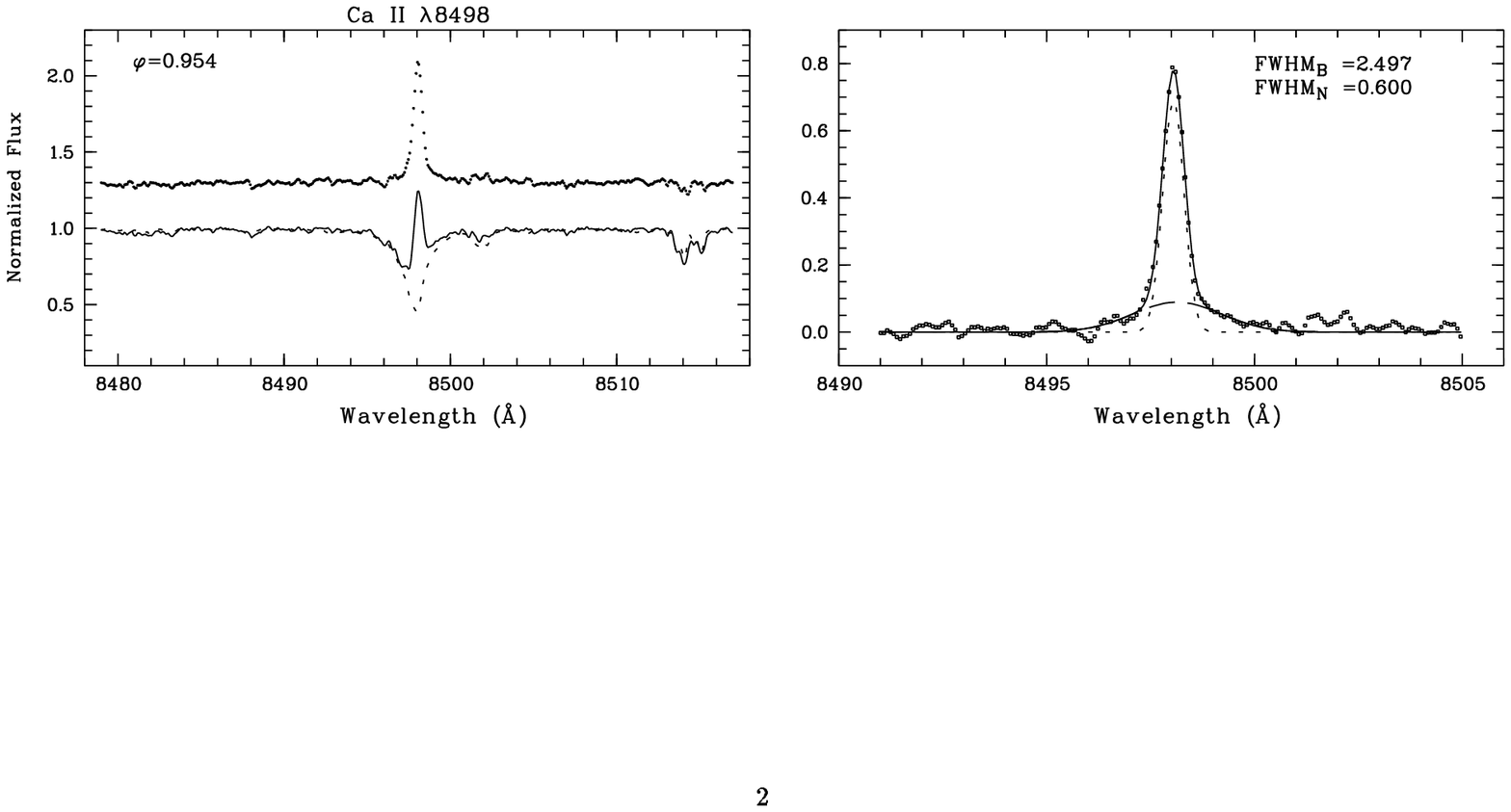,bbllx=28pt,bblly=284pt,bburx=585pt,bbury=448pt,height=5.6cm,width=18.0cm,clip=}}
\vspace{0.3cm}
{\psfig{figure=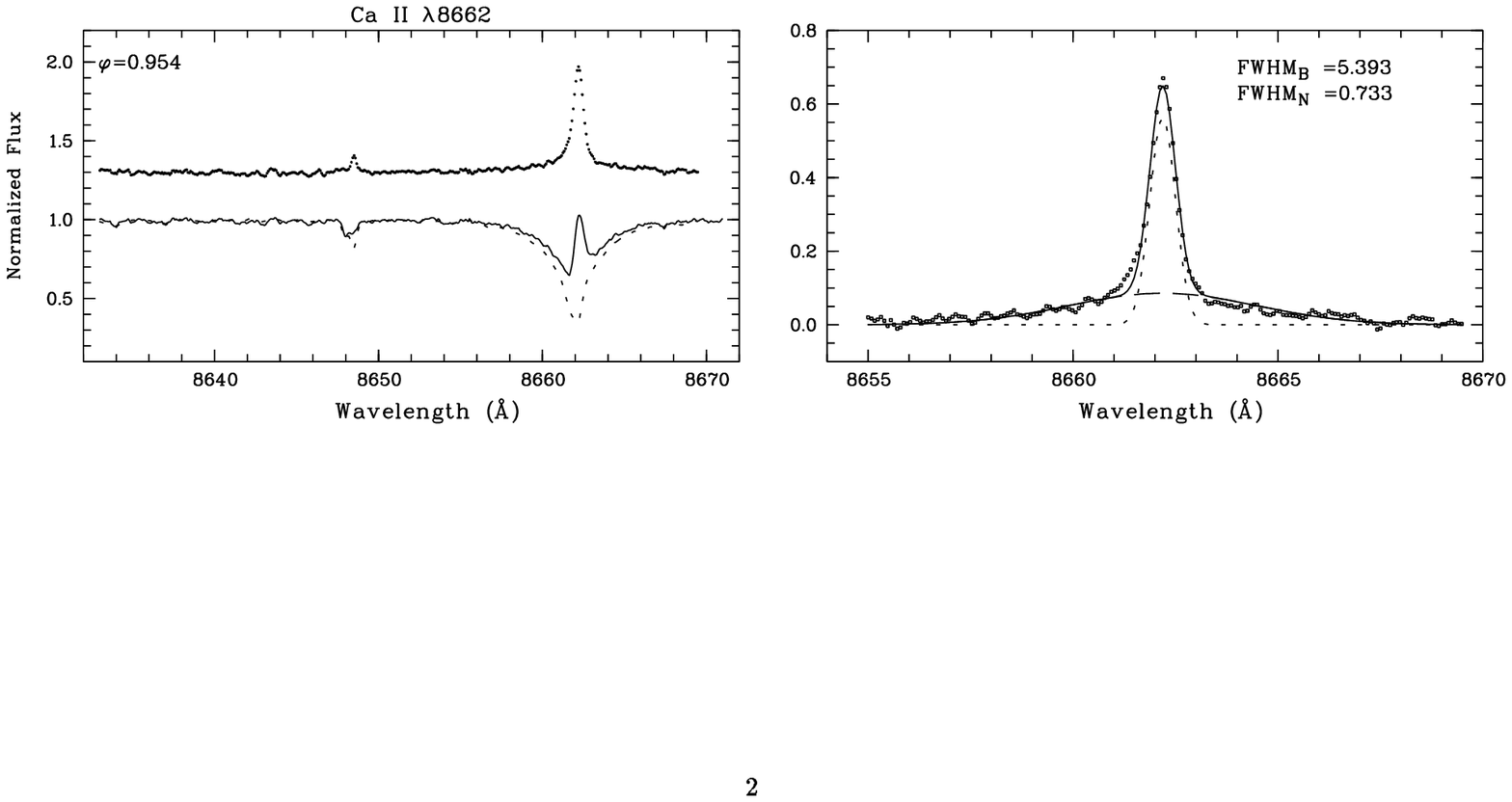,bbllx=28pt,bblly=284pt,bburx=585pt,bbury=448pt,height=5.6cm,width=18.0cm,clip=}}
\caption[ ]{WHT/UES Ca~{\sc ii} IRT $\lambda$ 8498 and
$\lambda$ 8662 spectra (Jul 1993), as in Fig~.\ref{fig:ues93ha}
\label{fig:ues93ca} }
\end{figure*}

\begin{figure*}
{\psfig{figure=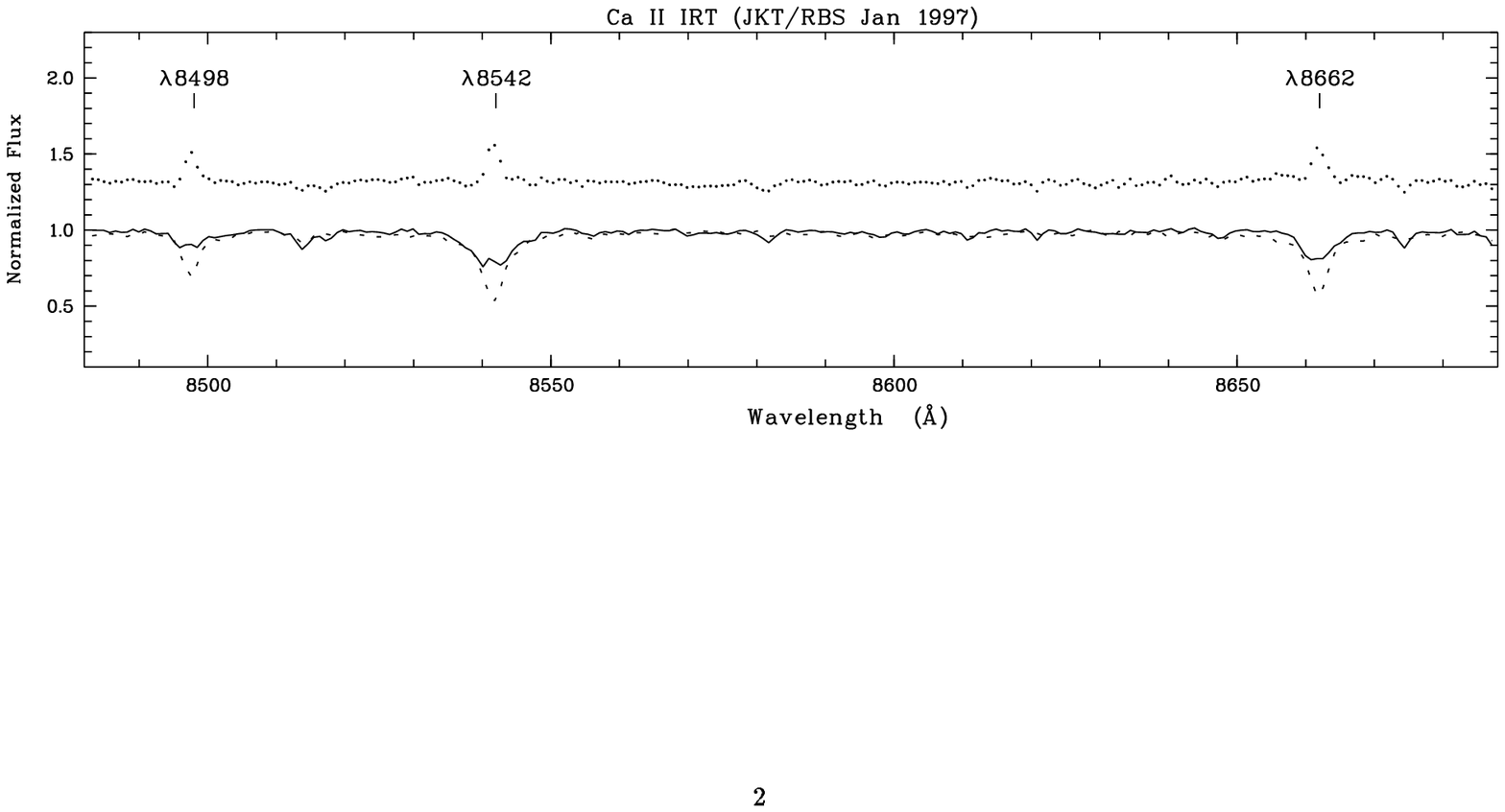,bbllx=28pt,bblly=284pt,bburx=585pt,bbury=448pt,height=5.6cm,width=18.0cm,clip=}}
\caption[ ]{JKT/RBS Ca~{\sc ii} IRT spectrum (Jan 1997) at orbital phase
0.511
\label{fig:jktcairt} }
\end{figure*}

\subsection{The Ca~{\sc ii} IRT lines}

The Ca~{\sc ii} infrared triplet (IRT) $\lambda$$8498$, $\lambda$$8542$, and 
$\lambda$$8662$ lines are other important chromospheric activity indicators
(Linsky et al. 1979; Foing et al 1989; Dempsey et al. 1993).
These lines share the upper levels of the Ca~{\sc ii} H \& K transitions
(see for example Mihalas 1978), but unlike the blue Ca~{\sc ii} lines, 
the IRT lies in a region of the spectrum with a well-defined continuum, 
making the calibration simpler. 
The IRT is formed in the lower chromosphere. 
Shine \& Linsky (1972, 1974) observed solar IRT plage profiles
concluding that the emission results from a steepening of the temperature
gradient in the chromosphere. This forces the temperature, electron density,
and Ca II density population to be higher for a given optical depth, 
compared to the quiet chromosphere, 
thus increasing the source function thereby producing
emission in the line cores.

In the observations carried out by Dempsey et al. (1993) only one star of
the sample (BY Dra) showed
emission above the local continuum in only the $\lambda$8498 line, and other 
seven stars revealed emission rising above the shallow line core.

The $\lambda$$8498$ and $\lambda$$8662$ lines are included in the 
1993 WHT/UES echelle spectrum of EZ Peg (see Fig.~\ref{fig:ues93ca}),
and both lines exhibit a strong emission superposed to the red wing of the 
corresponding absorption line, in agreement with the orbital phase (0.954).
In the 1997 JKT/RBS spectrum at orbital phase 0.511 
the three Ca~{\sc ii} IRT lines are included (see Fig.~\ref{fig:jktcairt}).
In this case the lines show a small emission reversal centered 
at the absorption line in the observed spectrum and a clear excess emission
in the difference spectrum.

Table~\ref{tab:measurescairt} gives
the Ca~{\sc ii} IRT line parameters, measured in the
observed spectra by reconstruction of the absorption line profile,
and using the spectral subtraction. The columns in this table
have the same meaning as for the Ca~{\sc ii} H \& K lines 
(see Table~\ref{tab:measureshyk}). 
Notice that the underestimation of 
the reconstruction method is larger for the IRT lines than for 
the Ca~{\sc ii} H \& K lines.
The absolute fluxes at the stellar surface have been obtained using  
the calibration of Hall (1996) as a function of (V~-~R).

As in the case of the H$\alpha$ line the subtracted profiles 
of the $\lambda$$8498$ and $\lambda$$8662$ lines also present
broad wings. Therefore we have used, as for the H$\alpha$ profile, 
a two gaussian (broad and narrow) fit.
 In Table~\ref{tab:measurescairt_nb} we list the parameters (I, FWHM, EW) 
of the broad and narrow components. 
The narrow components of both lines are 
very similar; however, the contribution of the broad component 
to the total EW is larger in the $\lambda$$8662$ line.
In the subtracted spectrum the intensity of the $\lambda$$8498$ line is 
higher than in $\lambda$$8662$; however, 
due to the contribution of the broad component, the total 
excess emission EW is larger in the $\lambda$$8662$ line.


\begin{table*}
\caption[]{Ca~{\sc ii} IRT lines 
measured in the observed and subtracted spectra  
\label{tab:measurescairt} }
\begin{flushleft}
\scriptsize
\begin{tabular}{l c c c c c c c c c c c c c}
\hline
\noalign{\smallskip}
 &     & \multicolumn{3}{c}{Reconstruction} &\ &
\multicolumn{3}{c}{Spectral subtraction} &\ &
\multicolumn{3}{c}{Absolute flux} \\
\cline{3-5}\cline{7-9}\cline{11-13}
\noalign{\smallskip}
\noalign{\smallskip}
 Obs.  & $\varphi$ &
EW & EW & EW & &
EW & EW & EW & &
$\log{\rm F}$ & $\log{\rm F}$ & $\log{\rm F}$\\
 & & 
$\lambda$$8498$ & $\lambda$$8542$ & $\lambda$$8662$ & & 
$\lambda$$8498$ & $\lambda$$8542$ & $\lambda$$8662$ & &
$\lambda$$8498$ & $\lambda$$8542$ & $\lambda$$8662$  \\
\noalign{\smallskip}
\hline
\noalign{\smallskip}
UES 1993  & 0.954 & 0.256 & - & 0.238 & & 0.680 & -     & 0.892 & &
 6.43 & -    & 6.55  \\
\noalign{\smallskip}
JKT 1997  & 0.511 & -     & - & -     & & 0.397 & 0.471 & 0.453 & &
 6.19 & 6.26 & 6.25  \\
\noalign{\smallskip}
\hline
\end{tabular}
\end{flushleft}
\end{table*}

\begin{table*}
\caption[]{Parameters of the broad and narrow Gaussian components
used in the fit of the Ca~{\sc ii} IRT subtracted spectra 
\label{tab:measurescairt_nb}}
\begin{flushleft}
\scriptsize
\begin{tabular}{lcccccccccc}
\noalign{\smallskip}
\hline
\noalign{\smallskip}
\noalign{\smallskip}
        &    &  
\multicolumn{4}{c}{Broad component} &\ &
\multicolumn{4}{c}{Narrow component} \\
\cline{3-6}\cline{8-11}
\noalign{\smallskip}
 {Obs.} & {$\varphi$} & 
{I} & FWHM & EW$_{\rm B}$ & EW$_{\rm B}$/EW$_{\rm T}$ & &  
{I} & FWHM & EW$_{\rm N}$ & EW$_{\rm N}$/EW$_{\rm T}$ \\
        &           & 
    & {\scriptsize (\AA)} & {\scriptsize (\AA)} & (\%) & & 
    & {\scriptsize (\AA)} & {\scriptsize (\AA)} & (\%) \\
\noalign{\smallskip}
\hline
\noalign{\smallskip}
Ca~{\sc ii} IRT line $\lambda$$8498$ \\
\noalign{\smallskip}
\hline
\noalign{\smallskip}
UES 1993 & 0.954 & 0.089 & 2.497 & 0.237 & 34.8 & & 
0.693 & 0.600 & 0.442 & 65.2 \\
\noalign{\smallskip}
\hline
\noalign{\smallskip}
Ca~{\sc ii} IRT line $\lambda$$8662$ \\ 
\noalign{\smallskip}
\hline
\noalign{\smallskip}
UES 1993 & 0.954 & 0.086 & 5.393 & 0.452 & 50.7 & & 
0.564 & 0.733 & 0.440 & 49.3 \\
\noalign{\smallskip}
\hline
\noalign{\smallskip}
\end{tabular}

\end{flushleft}
\end{table*}

\section{Conclusions}

In this paper we have analysed, using the spectral subtraction technique,
simultaneous spectroscopic observations of several optical
chromospheric activity indicators, at different epochs, 
of the RS CVn binary system EZ Peg.

We have found the H$\alpha$ line in emission above the continuum in all
the spectra, with noticeable changes in the line profile, that in part are due
to the superposition of the absorption line to the other component.
The subtracted H$\alpha$ profile is well matched using a two-components
Gaussian fit (narrow and broad).
The broad component is mainly responsible for the observed variations
of the subtracted profile, and its contribution to the total EW increases with
the degree of activity.
We have interpreted this broad component 
(also observed in other active systems in Paper I)
as arising from microflaring activity that take place
in the chromosphere. 

A filled-in in the H$\beta$ line has been found; however, 
the subtracted profiles do not present broad wings.
The ratio (E$_{\rm H\alpha}$/E$_{\rm H\beta}$) 
that we have found indicates that the emission would arise from extended 
regions viewed off the limb.

Strong Ca~{\sc ii} H \& K emission from the hot component 
is observed in our spectra.
The spectral subtraction reveals also a small but clear
emission in the H$\epsilon$ line.
The lines $\lambda$$8498$ and $\lambda$$8662$ of the  Ca~{\sc ii} IRT 
also show a strong emission.
As in the case of the H$\alpha$ line the subtracted profiles of these lines 
are well matched using a two-component
Gaussian fit.

We suggest that the He~{\sc i} D$_{3}$ could present a total filling-in 
due to microflaring. 
We have not found filled-in  by
chromospheric emission in the Na~{\sc i} D$_{1}$ and D$_{2}$ lines
nor in the  Mg~{\sc i} b triplet lines.

All the activity indicators analysed here indicate that the
hot component is the active star of the system,
contrary to the usual behaviour observed in 
chromospherically active binaries. 
The variations (in different epochs and with the orbital phase) 
observed in the different indicators, 
formed at different heights in the chromosphere, 
are correlated.

The analysis of several Ti~{\sc i} lines with a positive effect in 
luminosity class indicates that the hot component is not a main sequence
star but a subgiant or even of higher luminosity class,
in contrast with the cool component that seems to be a dwarf.
This result is confirmed by the application of the Wilson-Bappu effect to 
our Ca~{\sc ii} K spectra.
This difference in luminosity class, and therefore in 
convective zone depth, between the hot and cool components
could explain why the hot component is the active star of the system.


\begin{acknowledgements}

We thank N. Cardiel and J. Cenarro for having taken and reduced 
the 1997 JKT spectrum. 
This research has made use of La Palma Archive.
This work has been supported by the Universidad Complutense de Madrid
and the Spanish Direcci\'{o}n General de Investigaci\'{o}n
Cient\'{\i}fica y  T\'{e}cnica (DGICYT) under grant PB94-0263.

\end{acknowledgements}




\begin{thebibliography}{}

\bibitem{} Alden H.L., 1958, AJ, 63, 358

\bibitem{} Andretta V., Giampapa M.S., 1995, ApJ 439, 405

\bibitem{} Barrado D., Fern\'{a}ndez-Figueroa M.J., Garc\'{\i}a L\'{o}pez R.J.,
     De Castro, E., Cornide M., 1997, A\&A 326, 780 

\bibitem{} Barret P., 1996, PASP 108, 412

\bibitem{} Basri G., Wilcots E., Stout N., 1989, PASP 101

\bibitem{} Cardiel N., Gorgas J., 1997 (in preparation)

\bibitem{} Dempsey R.C., Bopp B.W., Henry G.W., Hall D.S., 1993, ApJS 86, 293

\bibitem{} Dempsey R.C., Neff J., Linsky J.L., Brown A., 1996a, IAU Symp. 176,
K.G. Strassmeier \& J.L. Linsky (eds.), Stellar Surface Structure, p. 411

\bibitem{} Dempsey R.C., Neff J., Thorpe M.J.,  et al.,
1996b, ApJ 470, 1172
          
\bibitem{} Favata F., Barbera M., Micela G., Sciortino S., 1993, A\&A 277, 428
                                
\bibitem{}  Fern\'andez-Figueroa M.J., Barrado D., De Castro E.,
Cornide M., 1993, A\&A 274, 373
         
\bibitem{}  Fern\'andez-Figueroa M.J., Montes D., De Castro E.,
Cornide M., 1994, ApJS 90, 433

\bibitem{} Foing B., Crivellari L., Vladilo G., Rebolo R., Beckman J., 1989,
A\&AS 80, 189

\bibitem{} Ginestet N., Carquillat J.M., Jashek M., Jashek C., 1994,
 A\&AS 108, 359

\bibitem{} Griffin R.F., 1985, The Observatory, 105, 81

\bibitem{} Gunn A.G., Hall J.C., Lockwood G.W., Doyle J.G., 1996,
A\&A 305, 146    

\bibitem{} Gunn A.G., Doyle J.G., 1997, A\&A 318, 60

\bibitem{} Gunn A.G., Doyle J.G., Houdebine E.R., 1997, A\&A 319, 211

\bibitem{} Hall J.C., Ramsey L.W., 1992, AJ 104, 1942

\bibitem{} Hall J.C., 1996, PASP 108, 313

\bibitem{} Howell S.B., Bopp B.W., 1985, PASP 97, 72

\bibitem{} Howell S.B., Williams W.M., Barden S.C., Bopp B.W.,
1986, PASP 98, 777    

\bibitem{} Huenemoerder D.P., Ramsey L.W., 1987, ApJ 319, 392

\bibitem{} Huenemoerder D.P., Ramsey L.W.,  Buzasi D.L., 1990, Cool star
stellar systems and the Sun, Sixth Cambridge Workshop., G. Wallerstein ed.,
ASP Conference Series 9, p 236

\bibitem{} Irvine N.J., 1972, PASP 84, 671

\bibitem{} Jashek C., Jashek M., 1995, "The Behavior of chemical elements in 
stars", Cambridge University Press
          
\bibitem{} Keenan P.C., Hynek J.A., 1945, ApJ 101, 265

\bibitem{} Kirkpatrick J.D., Henry T.J., McCarthy D.W., Jr., 1991,  ApJS 77, 
417
   
\bibitem{} Linsky J.L., Hunten D., Glacken D., Kelch W., 1979, ApJS 41, 481
    
\bibitem{} Linsky J.L., Wood B.E., 1994, ApJ 430, 342

\bibitem{} Linsky J.L., Wood B.E., Judge P., Brown A.,
Andrulis C., Ayres T.R., 1995, ApJ 442, 381

\bibitem{} Lutz T.E. 1970, AJ 75, 1007

\bibitem{} Mihalas D., 1978, in Stellar Atmospheres, 2d ed. (San Francisco: 
Freeman), 381

\bibitem{} Mitrou C.K., Doyle J.G., Mathioudakis M., Antonopoulou E.,
1996, A\&AS 115, 61
  
\bibitem{} Montes D., Fern\'{a}ndez-Figueroa M.J., De Castro E.,
 Cornide M., 1994, A\&A 285, 609


\bibitem{}  Montes D., Fern\'{a}ndez-Figueroa M.J., De Castro E.,
 Cornide M., 1995a, A\&A 294, 165

\bibitem{}  Montes D., Fern\'{a}ndez-Figueroa M.J., De Castro E.,
 Cornide M., 1995b, A\&AS 109, 135

\bibitem{}  Montes D., De Castro E., Fern\'{a}ndez-Figueroa M.J.,
 Cornide M. 1995c, A\&AS 114, 287

\bibitem{M1995d}  Montes D., Fern\'{a}ndez-Figueroa M.J., De Castro E.,
 Cornide M. 1995d, Stellar Surface Structure, IAU Symp 176,
Poster Proceedings, Strassmeier K. (ed), p. 167

\bibitem{M1996a}  Montes D.,  Fern\'{a}ndez-Figueroa M.J.,
 Cornide M., De Castro E., 1996a, A\&A 312, 221

\bibitem{M1996b} Montes D., Sanz-Forcada J., Fern\'{a}ndez-Figueroa M.J.,
Lorente R., 1996b, A\&A 310, L29

\bibitem{M1997} Montes D., Mart\'{\i}n E.L., Fern\'{a}ndez-Figueroa M.J.,
Cornide M., De Castro E. 1997a, A\&AS 123, 473 

\bibitem{M1997} Montes D., Fern\'{a}ndez-Figueroa M.J., De Castro E.,
Sanz-Forcada J., 1997b, A\&AS 125 (in press) (Paper I)

\bibitem{} Pallavicini R., Cerruti-Sola M., Duncan D.K., 
1987, A\&A 174, 116

\bibitem{} Pasquini L., Pallavicini R., Pakull M., 1988,
 { A\&A} {  191}, 253

\bibitem{} Pasquini L., Pallavicini R. 1991, A\&A 251, 199

\bibitem{} Robinson R.D., Airapetian V.S., Maran S.P., Carpenter K.G.,
1996, ApJ 469, 872

\bibitem{} Saar S.H., Huovelin J., Osten R.A., Shcherbakow A.G., 1997, A\&A
326, 741

\bibitem{} Schlesinger F., Barney I., Gesler C., 1934, Yale Trans., 10, 170

\bibitem{} Schmidt-Kaler T. 1982, {  in Landolt-B\"{o}rnstein,} Vol.{  2b},
 ed K. Schaifers, H.H. Voig (Heidelberg: Springer)

\bibitem{} Shine R.A., Linsky J.L., 1972, SoPh, 25, 357

\bibitem{} Shine R.A., Linsky J.L., 1974, SoPh, 39, 49 

\bibitem{} Soderblom D.R., Oey M.S., Johnson D.R.H., Stone R.P.S., 
1990, AJ 99, 595

\bibitem{} Strassmeier K.G., Hall D.S., Fekel F.C., Scheck M.,
 1993, { A\&AS} {\ 100}, 173 (CABS)

\bibitem{} Szkody P., 1977, ApJ 217, 140

\bibitem{} Szkody P., Michalsky J.J., Stokes G.M., 1982, PASP 94, 137  

\bibitem{} Vilkki E.U., Welty D.E., Cudworth K.M., 1986, AJ 92, 989 

\bibitem{} Vyssotsky A.N., Balz A.G.A., 1958, Leander McCormick Publ., 
13, 79 

\bibitem{} Wilson O.C., Bappu M.K.V. 1957, { ApJ} {  125},
661

\bibitem{} Wood B.E., Harper G.M., Linsky J.L., Dempsey R.C., 1996,
ApJ 458, 761

\bibitem{} Wood B.E., Linsky J.L., Ayres T.R., 1997, ApJ 478, 745

\bibitem{ } Zuiderwijk E.J., Martin R., Raimond E., van Diepem G.N.J.
1994, PASP 106, 515.
                                       
\end{thebibliography}
\end{document}